%% file: main.tex
\documentclass[sigconf]{acmart}

\usepackage{hyperref}
\usepackage{url}
\usepackage{breakurl}
\hypersetup{
   breaklinks=true,   % splits links across lines
   colorlinks=true,   % displays links as colored text instead of blocks
   pdfusetitle=true,  % \title and \author values into pdf metadata
}

\usepackage{amsmath}
\usepackage{graphicx}
\usepackage{tikz}
\usepackage{algorithm}
\usepackage{algpseudocode}
\usepackage{caption}
\usepackage{subcaption}
\usepackage{lipsum}
\usepackage{multicol}
\usepackage{soul}
\usepackage[frozencache,cachedir=.]{minted}

\newcommand{\revision}[1]{{\color{black}#1}}

\definecolor{verbgray}{gray}{0.9}

\usepackage{lipsum}
\usepackage{xcolor}
\usepackage{tcolorbox}
\tcbuselibrary{breakable}
\usepackage{lipsum}
\usepackage{colortbl}
\usepackage{ragged2e}
\usepackage{blindtext}
\usepackage{enumitem}
\setlist{leftmargin=5.5mm}

\newtcolorbox{eqbox}[1][]
{
colframe=gray!40,
colback=gray!10,
coltitle=gray!20!black,
arc=0pt,
boxrule=0pt,
left=3pt,right=3pt,top=2pt,bottom=2pt,
title=#1
}

\AtBeginDocument{%
  }

\copyrightyear{2024} 
\acmYear{2024} 
\setcopyright{rightsretained} 
\acmConference[CHI '24]{Proceedings of the CHI Conference on Human Factors in Computing Systems}{May 11--16, 2024}{Honolulu, HI, USA}
\acmBooktitle{Proceedings of the CHI Conference on Human Factors in Computing Systems (CHI '24), May 11--16, 2024, Honolulu, HI, USA}
\acmDOI{10.1145/3613904.3642596}
\acmISBN{979-8-4007-0330-0/24/05}

\begin{document}

\title[Is Stack Overflow Obsolete? An Empirical Study of ChatGPT Answers to Stack Overflow Questions]{Is Stack Overflow Obsolete? An Empirical Study of the Characteristics of ChatGPT Answers to Stack Overflow Questions}
% arxiv title
% \title{Who Answers It Better? An In-Depth Analysis of ChatGPT and Stack Overflow Answers to Software Engineering Questions }

% \affiliation{%

\author{Samia Kabir}
\affiliation{%
  \institution{Purdue University }
  % \streetaddress{1 Th{\o}rv{\"a}ld Circle}
  \city{West Lafayette}
  \country{USA}}
% \email{kabirs@purdue.edu}

\author{David N. Udo-Imeh}
\affiliation{%
  \institution{Purdue University }
  % \streetaddress{1 Th{\o}rv{\"a}ld Circle}
  \city{West Lafayette}
  \country{USA}}
% \email{dudoimeh@purdue.edu}

\author{Bonan Kou}
\affiliation{%
  \institution{Purdue University }
  % \streetaddress{1 Th{\o}rv{\"a}ld Circle}
  \city{West Lafayette}
  \country{USA}}
% \email{koub@purdue.edu}

\author{Tianyi Zhang}
\affiliation{%
  \institution{Purdue University }
  % \streetaddress{1 Th{\o}rv{\"a}ld Circle}
  \city{West Lafayette}
  \country{USA}}
\begin{abstract}
    Q\&A platforms have been crucial for the online help-seeking behavior of programmers. However, the recent popularity of ChatGPT is altering this trend. Despite this popularity, no comprehensive study has been conducted to evaluate the characteristics of ChatGPT’s answers to programming questions. To bridge the gap, we conducted the first in-depth analysis of ChatGPT answers to 517 programming questions on Stack Overflow and examined the correctness, consistency, comprehensiveness, and conciseness of ChatGPT answers. Furthermore, we conducted a large-scale linguistic analysis, as well as a user study, to understand the characteristics of ChatGPT answers from linguistic and human aspects. Our analysis shows that 52\% of ChatGPT answers contain incorrect information and 77\% are verbose. Nonetheless, our user study participants still preferred ChatGPT answers 35\% of the time due to their comprehensiveness and well-articulated language style. However, they also overlooked the misinformation in the ChatGPT answers 39\% of the time.
    % \todo{The numbers are not consistent with the number in intro.}.
    This implies the need to counter misinformation in ChatGPT answers to programming questions and raise awareness of the risks associated with seemingly correct answers. 
\end{abstract}




\maketitle

\input{sections/Intro}

\input{sections/Relatedwork}

% \input{sections/todos}
\input{sections/RQ}

\input{sections/Methodology}
% \input{sections/Opencoding}
% % \input{sections/Data Collection}
% % \input{sections/Analysis}
% \input{sections/Linguistic}
% \input{sections/Humaneval}
\input{sections/Results_manual}

\input{sections/Result_linguistic}
\input{sections/Result_userstudy}
\input{sections/Discussion}

\input{sections/Threat}

\input{sections/Conclusion}

\bibliographystyle{ACM-Reference-Format}
\bibliography{ref}

\input{appendix}
\end{document}

%% file: sections/Intro.tex
\section{Introduction}

%In the contemporary age of rapid software evolution, software developers frequently turn to web-based assistance.
Programmers often resort to online resources for a variety of programming tasks, e.g., API learning, bug fixing, comprehension of code or concepts, etc.~\cite{skripchuk2023analysis,xia2017developers,rao2020analyzing}.
% They employ various approaches, such as utilizing search engines to locate documentation, code samples, tutorials, or blog posts.
A vast majority of these help-seeking activities include frequent engagement with community Q\&A platforms such as Stack Overflow %\footnote{https://stackoverflow.com/}
(SO)~\cite{rao2020analyzing,rahman2018evaluating,xia2017developers,vasilescu2013stackoverflow}. The emergence of \textit{Large Language Models (LLMs)} has demonstrated the potential to transform the online help-seeking patterns of programmers. 
%Recent studies show that programmers utilize AI tools such as GitHub Copilot~\cite{copilot} for rapid suggestions and only turn to web searches when they need to verify a solution or access the documentation~\cite{barke2023grounded,ross2023programmer,vaithilingam2022expectation}. 
% The ability to engage in \todo{GitHub Copilot is not a conversational agent. It's more of an auto-completion tool. ChatGPT is conversational.}interactive conversations and provide apt solutions using natural language has propelled \textit{LLM}s into becoming a popular option among programmers.
% Moreover, \textit{LLM}s have demonstrated commendable performance across various downstream software engineering tasks ~~\cite{ross2023programmer} such as code translation between different programming languages~~\cite{lachaux2020unsupervised}, code auto-completion~~\cite{copilot}, code generation based on natural language descriptions~~\cite{xu2021ide}, etc. 
In November 2022, ChatGPT~\cite{chatgpt} was released and quickly gained significant attention and popularity among programmers. 
% The popularity of ChatGPT can be attributed to several key factors.ChatGPT's capacity to engage in human-like conversations and promptly learn from continuous human feedback have contributed to its popularity. 
There have been increasing debates about whether and when ChatGPT would replace prominent search engines and Q\&A forums among researchers and industrial practitioners~\cite{tweet, quora}.

Despite the rising popularity of ChatGPT, there are also many increasing concerns. Previous studies show that \textit{LLM}s can acquire factually incorrect knowledge during training and propagate the incorrect knowledge to generated content~\cite{goodrich2019assessing,maynez2020faithfulness,gamage2022deepfakes, gehman2020realtoxicityprompts, bender2021dangers}. Besides, \textit{LLM}s often generate fabricated texts that mimic truthful information and are hard to recognize, especially for users who lack the expertise~\cite{cao2021knowledgeable,bommasani2021opportunities,elazar2021measuring}. Like other \textit{LLM}s, ChatGPT is also plagued with these issues~\cite{guo2023close,borji2023categorical,mitrovic2023chatgpt,kocon2023chatgpt}. The prevalence of misinformation, which can easily mislead users, has prompted Stack Overflow to impose a ban on answers generated by ChatGPT~\cite{soban}.

Recent studies have compared ChatGPT to human experts in legal, medical, and financial domains~\cite{guo2023close, gao2022comparing}. To the best of our knowledge, no comprehensive analysis has been conducted to investigate ChatGPT's capability to answer programming questions, especially the quality and characteristics of ChatGPT answers in comparison to human answers. If misinformation is prevalent in ChatGPT answers and is hard to recognize, it may inevitably lead to suboptimal design choices and software defects. In the long term, this may jeopardize the quality and robustness of software and cyberinfrastructure in our society, affecting a broader population beyond programmers.   
% \todo{rephrase, no study to analyze the characteristics, factor to contribute to Chatgpt's quality, }
%Moreover, there is no investigation into the factors contributing to the quality and characteristics of ChatGPT answers to programming questions. 
% Moreover, no formal study has been conducted investigating the factors of the popularity of ChatGPT. Hence, the question of whether ChatGPT can potentially replace popular Q\&A platforms such as Stack Overflow remains unexplored.

This work aims to bridge the gap by adopting a mixed-methods research design~\cite{johnson2004mixed} with a combination of manual analysis, linguistic analysis, and user studies to compare human answers and ChatGPT answers to programming questions on Stack Overflow (SO). Specifically, we performed stratified sampling to collect ChatGPT answers to 517 SO questions with different characteristics (e.g., popularity, question types, recency, etc.). The sample size is statistically significant with a 95\% confidence level and 5\% margin of error. We manually analyzed ChatGPT answers and compared them with the accepted SO answers written by human programmers. %through Open Coding~\cite{hancock2001introduction}. 
In addition to correctness, we also assessed the consistency, comprehensiveness, and conciseness of ChatGPT answers. We found that 52\% of ChatGPT answers contain misinformation, 77\% of the answers are more verbose than human answers, and 78\% of the answers suffer from different degrees of inconsistency to human answers.

Furthermore, to examine how the linguistic features of ChatGPT answers differ from human answers, we conducted a large-scale linguistic analysis on 2000 randomly sampled SO questions. Specifically, we run Linguistic Inquiry and Word Count (LIWC)~\cite{pennebaker2015development} and sentiment analysis on ChatGPT answers and human answers. Our results show that ChatGPT uses more formal and analytical language and portrays less negative sentiment.

Finally, to capture how different characteristics of the answers influence programmers' preferences between ChatGPT and SO, we conducted a user study with 12 programmers. The study results show that participants' overall preferences, correctness ratings, and quality ratings were more leaning toward human answers from Stack Overflow. However, participants still preferred ChatGPT answers 35\% of the time and overlooked misinformation in the answers 39\% of the time.
% \todo{numbers not consistent with abstract. 39\% for preference or overlooking misinformation} \todo{You also need to add a sentence to describe how often participants overlooked misinformation} \new{done}
%despite the incorrectness of ChatGPT answers. %Nonetheless, they successfully differentiated between machine and human-generated answers about 80\% of the time. 
When asked why they preferred ChatGPT answers even when they were incorrect, participants suggested the comprehensiveness and articulated language structures of the answers as reasons for their preference, which is consistent with our linguistic analysis result.

Our manual analysis, linguistic analysis, and user study collectively demonstrate that while ChatGPT performs remarkably well in many cases, it frequently makes errors and unnecessarily prolongs its responses. However, ChatGPT answers have richer linguistic features, leading some users to prefer ChatGPT answers over human answers and sometimes overlook the underlying incorrectness and inconsistencies in ChatGPT answers. Our in-depth analysis points towards several challenges and risks of using ChatGPT in programming and also highlights several opportunities for designing new interaction and computational methods to counter misinformation generated by ChatGPT. %the induced ignorance and risks that are associated with the significant number of incorrectness. 

To conclude, this paper makes the following contributions:
\begin{itemize}
    \item We conducted an in-depth analysis of the correctness and quality of ChatGPT answers across four distinct quality aspects for various types of SO question posts. 

    \item We performed a large-scale analysis of the linguistic characteristics of ChatGPT answers and identified distinct linguistic features that are prominent in ChatGPT answers. 

    \item We investigated how real programmers consider answer correctness, quality, and linguistic features when choosing between ChatGPT and Stack Overflow (SO) through a within-subjects user study. 
    
    \item We provided a comprehensive discussion of the design implications, emphasized the risks of misinformation, and outlined future directions aimed at detecting and mitigating misinformation in AI-assisted programming.
    
    \item{We made our data and codebooks publicly available at 
    % \href{https://github.com/SamiaKabir/ChatGPT-Answers-to-SO-questions}
    \burl{https://github.com/SamiaKabir/ChatGPT-Answers-to-SO-questions} to foster future research in this direction.
    % \footnote{We use an anonymous repository for now to follow the double-blind review policy at CHI.}
    }
\end{itemize}

The rest of the paper is organized as follows. Section 2 describes the related work. Section 3 describes the research questions. Section 4 describes the data collection process and the methodology of our mixed-methods study.
 % -- 3.2 for manual analysis, 3.3 for linguistic analysis, and 3.4 for user study.
 Sections 5, 6, and 7 describe the analysis results of our manual analysis, linguistic analysis, and user study respectively. Section 8 discusses the implications of our findings and future research directions. Section 9 describes the limitations of this work. Section 10 concludes this work.

%% file: sections/Relatedwork.tex
\section{Related Work}

\subsection{Misinformation Generated by LLMs}
Previous studies have shown that content generated by \textit{LLM}s may contain hallucinations and misinformation~\cite{bommasani2021opportunities,westerlund2019emergence,gptdupe}. Some recent work has investigated \textit{LLM}s' capability to generate fake news, images, and videos~\cite{zhou2023synthetic, textdeepfake, gehman2020realtoxicityprompts} about a multitude of social phenomena such as politics, elections, people, disease, economics, etc. 
These types of misinformation have the power to mislead and misguide people and can potentially impair the normal functionalities of society and cause chaos~\cite{textdeepfake2,gamage2022deepfakes,islam2020covid, gptdupe}. Specifically, several studies have investigated the power of AI-generated texts in deceiving people~\cite{kreps2022all,buchanan2021truth}.  Zhou et al.~\cite{zhou2023synthetic} further highlight the risks by showing how traditional misinformation detection and mitigation methods often fail to identify misinformation generated by state-of-the-art \textit{LLM}s.

Since its release in November 2022, ChatGPT has surpassed other \textit{LLM}s in popularity among general users. The usability and effectiveness of ChatGPT have been examined in different domains, such as law, medicine, and finance~\cite{guo2023close}. Like other \textit{LLM}s, ChatGPT also fabricates facts and generates low-quality or misleading information~\cite{guo2023close,borji2023categorical,mitrovic2023chatgpt,kocon2023chatgpt}. 
However, to the best of our knowledge, no studies have investigated the characteristics and human perception of misinformation in ChatGPT's answers to programming questions. Our work aims to bridge this gap with a combination of manual analysis, linguistic analysis, and user studies.

\subsection{Help-Seeking Behavior of Programmers}
The proliferation of social media and Q\&A platforms for programming have immensely shaped the online help-seeking behavior of programmers~\cite{treude2011programmers,storey2010impact, vasilescu2013stackoverflow, abdalkareem2017developers}. %Existing literature investigates the role and benefits of Q\&A platforms~\cite{treude2011programmers,mamykina2011design,parnin2012crowd}. 
Treude et al.~\cite{treude2011programmers} investigated the role and benefits of a popular Q\&A platform---Stack Overflow (SO)---and found that Stack Overflow is highly effective in code reviews and answering conceptual questions. Through a mixed-methods study, Mamykina et al.~\cite{mamykina2011design} show that the chance of getting quick answers from the SO community is high. %, with a median time of 11 minutes. 
% which also contributes towards the wide range acceptance of SO. 
Despite the popularity and effectiveness of Stack Overflow, several concerns have been raised. %that Stack Overflow may interrupt developers' workflow and impair their performance persistence~\cite{vasilescu2013stackoverflow,storey2010impact}.
For instance, since Q\&A platforms are not integrated with IDEs, developers have to constantly switch between their IDEs and Q\&A platforms, which may interrupt developers' workflow and impair their performance persistence~\cite{vasilescu2013stackoverflow,storey2010impact, bacchelli2012harnessing}. 
Another concern is the presence of toxicity and negative sentiment in people's answers and comments on Stack Overflow. Calefato et al.~\cite{calefato2015mining} found that the presence of positive and negative sentiment contributes towards the upvotes and downvotes of SO answers respectively. In a follow-up study, they found that novices and student programmers often encounter arrogant and rude comments on Stack Overflow, which discourages them from posting questions~\cite{calefato2018ask}. Asaduzzaman et al.~\cite{asaduzzaman2013answering} also found that the presence of toxicity and negative emotions in SO answers can discourage follow-up discussions on Stack Overflow.
Our study also confirms this, since users prefer ChatGPT answers due to their politeness and positive sentiment.

\subsection{Human-AI Collaboration in Programming}
%\todo{The first sentence is not correct. There have been many AI-based code completion tools like TabNine and VS Code Intelligence. Please rewrite.}The emergence of GitHub Copilot~\cite{copilot} addresses the limitation of non-integrated assistive AI programming tools by integrating them in IDEs in the form of code auto-completion. 
%Compared with prior AI-based code completion tools,\todo{add citations to these tools} Copilot can provide further assistance such as code translation, explanations, and documentation generation~\cite{ross2023programmer,copilotlab}. 
Recent studies show that AI pair-programming tools such as GitHub Copilot~\cite{copilot} have shifted developers' behavior from code writing to code understanding and can improve developer productivity~\cite{bird2022taking, imai2022github}. The online help-seeking behavior of developers is also changing along with other behavior shifts. Developers often use Copilot to get quick code suggestions and only turn to web searches to access the documentation or verify the suggestions~\cite{barke2023grounded,ross2023programmer,vaithilingam2022expectation}. However, recent studies show that Copilot often generates code with errors, which can become a liability for programmers~\cite{dakhel2023github}. Furthermore, programmers also need to debug and fix those errors and make other modifications in order to integrate generated code into their program context, which may, in turn, impair their productivity~\cite{vaithilingam2022expectation}. %Vaithilingam et al.~\cite{vaithilingam2022expectation} show that despite the limitations, developers still prefer using Copilot as a convenient starting point. 

ChatGPT has gained popularity among programmers of all levels since its release in November 2022.
% \todo{when I read here, I realize that reviewers may ask you one question---isn't ChatGPT integrated into Copilot? Instead of saying ChatGPT has gained popularity since 2022 and giving a direct comparison between ChatGPT and Copilot, can we jut say that ChatGPT has been integrated into Copilot and makes it conversational?}
One of the main advantages of ChatGPT over GitHub Copilot is that ChatGPT works as a conversational chatbot that allows users to ask questions and give feedback beyond code completion. For instance, programmers can ask a conceptual question about a data type used in a program, ask for a code explanation, and ask how to fix an error message~\cite{copilotvsgpt}. Recently, GitHub announced GitHub Copilot X, which integrates GPT-4, a more advanced version of the \textit{LLM} behind ChatGPT, into Copilot~\cite{copilotx,copilotx2}. 
% \todo{This sounds repetitive to the ending of Section 2.1.}
Despite the popularity of ChatGPT among programmers, to the best of our knowledge, there is still no in-depth and comprehensive analysis of the characteristics and quality of ChatGPT answers to programming questions. We bridge this research gap by empirically studying ChatGPT answers to programming questions on Stack Overflow.

% A very recent study by Xu et al.~\cite{xu2023we} compared ChatGPT answers with SO answers via automatic and manual comparison. Their results show that ChatGPT answers are semantically similar to human answers, but has low-quality compared to human answers.

%% file: sections/RQ.tex
\section{Research Questions}
\label{sec:rq}
This section describes the research questions investigated in this work and the rationale of each research question. The findings of these research questions will deepen our understanding of the characteristics of and the human perception of ChatGPT answers. They will also shed light on the challenges and risks of using ChatGPT for programming and inform the design of new interactive and computational methods to counter misinformation generated by ChatGPT. 
% The main objective of this work is to study different characteristics of ChatGPT answers in comparison with human-written answers on Stack Overflow. This investigation will help us understand the 
% characteristics of responses generated by state-of-the-art conversational models such as ChatGPT in the light of programming. Moreover, this work will help deepen our understanding of the human perception of correctness or quality issues of ChatGPT-generated solutions for programming tasks. We believe these findings can shed light on the challenges and risks of using ChatGPT for programming and inform the design of new interactive and computational methods to counter misinformation generated by ChatGPT. 
% % We aim to inspect and evaluate the ChatGPT answers from different aspects and identify the fine-grain issues of ChatGPT answers w.r.t human-generated answers (answered in SO).
% % % , through manually analyzing ChatGPT answers. 
% % We also aim to study the differences in terms of language features and identify the features that are inherent to ChatGPT-generated answers only.
% % % via linguistic analysis. 
% % % We expect to identify language features that are inherent to ChatGPT-generated answers only. 
% % And finally, want to study the user behavior and preference towards ChatGPT answers compared to human-generated answers.
% In particular, We investigated the following research questions and provided the motivation for each question below.

% \begin{FlushLeft}

\begin{itemize}

 \item {\em RQ1. How do ChatGPT answers differ from SO answers in terms of correctness and quality?}     Previous work~\cite{cao2021knowledgeable,huang2021factual,kocon2023chatgpt} has shown that \textit{LLM}s such as ChatGPT are prone to hallucination and may generate content with low quality. Therefore, we want to assess and quantify the correctness and different quality aspects (e.g., consistency, conciseness, comprehensiveness) of ChatGPT answers to programming questions. %We expect that with manual analysis, we can empirically study the correctness and quality factors of ChatGPT answers. 
    % By taking several characteristics of answer quality into consideration, we can report a comprehensive list of differences between ChatGPT and SO answers. 
    
% \tcblowerL
 \item {\em RQ2. What are the fine-grained issues associated with each of the correctness and quality aspects?}
 While RQ1 aims to provide a quantification of the correctness and quality of ChatGPT answers, RQ2 aims to conduct an in-depth, qualitative analysis and develop a taxonomy of the issues in ChatGPT answers. For instance, we are interested in finding out the common symptoms of hallucinations, e.g., conceptual errors, code errors, terminology errors, etc. %For example, some parts of an answer can be factually incorrect, some parts can have logical errors, or some parts can be irrelevant or redundant. %To answer this RQ, we conduct an in-depth analysis of ChatGPT answers to identify these fine-grained quality issues in ChatGPT answers.
 % By manually analyzing and annotating every sentence of an answer, we expect to find finer details and issues concerning the ChatGPT answers.

% \tcblower
   \item   {\em RQ3. Do the types of SO questions affect the quality of ChatGPT answers?} 
   Previous studies~\cite{allamanis2013and, kou2023automated} show that linguistic forms of human answers on Stack Overflow vary based on the types of programming questions. For example,  \textit{How-to} questions have step-by-step answers, while \textit{conceptual} questions contain descriptions and definitions. We seek to understand if the types of programming questions influence the characteristics of ChatGPT answers in a similar manner. 
   % We seek to find if the types of SO questions have any effect on the ChatGPT answer quality. For example, between answers to the questions that were posted after ChatGPT was released compared to questions posted before the release, does the quality differ? 

    \item  {\em RQ4. Do the language structure and attributes of ChatGPT answers differ from SO answers?} Previous studies~\cite{zhou2023synthetic} show that human-crafted misinformation and machine-generated misinformation have distinct linguistic features, which can facilitate misinformation detection. Prior work~\cite{bazelli2013personality} has also shown a relationship between linguistic characteristics and the acceptance of Stack Overflow answers. Inspired by these findings, we want to investigate the distinct linguistic characteristics of ChatGPT answers and how they compare to accepted SO answers written by human programmers. 
    % Identifying linguistic features that are distinctive to ChatGPT answers is an important way to understand if or how answers' certain linguistic features guide user preference. 

   \item   {\em RQ5. Do the underlying sentiment of ChatGPT answers differ from SO answers?} Previous studies~\cite{miller2022did,qiu2019signals} discuss the harmful effect of toxicity or negative tone in online discussions. Prior work~\cite{calefato2015mining} also shows the role of underlying sentiment in the acceptance of SO answers. Therefore, we seek to analyze the sentiment of ChatGPT answers and compare it to accepted answers on Stack Overflow. 
   % We want to analyze how the underlying sentiment of ChatGPT answers differs from SO. 
   % We are especially interested to see if the underlying tone of ChatGPT answers can be a contribution towards its' growing popularity. 

  % \item    \textbf{RQ6.} Do the distinctive expressions ChatGPT uses differ from that of human experts? 

  \item  {\em RQ6. Can programmers differentiate ChatGPT answers from human answers?}
    We are curious about whether programmers can discern machine-generated answers from human-written answers and what kinds of heuristics they employ to make the decision.
    % With a user study, we can understand if users can readily differentiate between machine and human-generated answers when they are unaware of the source of the answer. 
    %We seek to identify the distinct characteristics of machine-generated answers that make them distinguishable. 
    Investigating these heuristics is important since it helps identify good practices that can be adopted by the programmer community and inform the design of automated mechanisms.  %We also aim to determine if these findings align with findings from manual and linguistic analysis. 

   \item   {\em RQ7. Can programmers identify misinformation in ChatGPT answers?}
 Understanding how programmers identify misinformation in ChatGPT answers is important as it can provide insights about effective mechanisms to counter misinformation. %In this study, we aim to evaluate if a user can determine the misinformation in ChatGPT answers.  
 If programmers can identify the misinformation properly, we expect to find out the techniques of identification. Otherwise, if programmers struggle to identify misinformation, we expect to find out the challenges.
 % By designing a user study in a way that consists of some answers where ChatGPT is incorrect,  

  \item    {\em RQ8. Do programmers prefer ChatGPT over Stack Overflow?}
Finally, we want to understand the user preference between ChatGPT and human-generated answers based on the correctness, quality, and linguistic characteristics of the answer. 
% By hiding the source of the answer from the users, we expect to study how the user chose between human and machine-generated answers and what explicit characteristics of the answers prompt users to prefer one answer over the other. 

  % \item    \textbf{RQ9. Do users need additional support with ChatGPT answers compared to SO? \todo{discard it may be and discuss it inside the result of RQ8?}}

\end{itemize}
% \end{FlushLeft}
% \vspace{-10pt}

%% file: sections/Methodology.tex
\section{Methodology}

We adopted a mixed-methods research design to answer the research questions in Section \ref{sec:rq}. Specifically, to answer RQ1-RQ3, we conducted an in-depth manual analysis through open coding and thematic analysis (Section \ref{sec:opencode}). To answer RQ4 and RQ5, we conducted a large-scale linguistic analysis and sentiment analysis using automated methods (Section \ref{sec:linguistic}). To address RQ6 to RQ8, we conducted user studies followed by semi-structured interviews with 12 participants  
 (Section \ref{sec:userstudy}). The following sections provide a detailed description of each method.

% \begin{figure*}[!htp]
%     \centering
%     \includegraphics[width=1.0\textwidth]{figures/overview.png}
%     \caption{Overview of data collection and analysis process}
%     \label{fig:overview}
% \end{figure*}

\input{sections/DataCollection}

\input{sections/Opencoding}
\input{sections/Linguistic}

\input{sections/userstudy}

%% file: sections/DataCollection.tex
\subsection{Data Collection}

\subsubsection{\textbf{SO Question Collection}}
\label{datacol}

We consider three characteristics of programming questions---question popularity, posting time, and question type. %By manually observing the questions, we selected three categories-- \textit{Question Type, Question Popularity,} and \textit{Question Recency.} 
%To ensure that the human-written answers have good quality, we only consider SO questions that have an accepted answer post. 
We adopted a stratified sampling strategy to collect a balanced set of SO questions that fall into different categories w.r.t.~their popularity, posting time, and question type. Table \ref{tab:questiondist} shows the distribution of the sampled questions. We describe the sampling procedure below.

First, we collected all questions in the SO data dump (March 2023)~\cite{sodatadump} and ranked them by their view counts. We used view counts as the popularity metric of SO questions. We selected three categories of questions---the top 10\% of questions in the view count ranking ({\em Highly Popular}), the questions in the middle ({\em Average Popular}), and the bottom 10\% in the ranking ({\em Unpopular}). 

Second, from the three categories of questions above, we moved on to categorize them by their recency.  We split questions in each popularity category into two recency categories---questions posted before the release of ChatGPT (November 30, 2022) as \textit{Old}, and questions posted after that time as \textit{New}. We selected the release date of ChatGPT to evaluate how the answer characteristics of ChatGPT reflect the presence or absence of specific knowledge in ChatGPT's training data.

Third, for question types, based on the literature \cite{treude2011programmers,allamanis2013and,de2014ranking,kou2022sosum}, we focused on three common question types---\textit{Conceptual, How-to}, and \textit{Debugging}. 
% Following the same process as Kou et al.~\cite{kou2023automated},
We followed prior work~\cite{iyer2016summarizing, kou2023automated} and trained a Support Vector Machine (SVM) classifier to predict the type of a SO question based on the question title. The classifier achieves an accuracy of 78\%, which is comparable to prior work. Then, we used this classifier to predict the question type of SO questions in each category of questions obtained from the two previous steps. 

In the end, we randomly sampled the same number of questions from each category along the three aspects. Given that the question type classifier may not be accurate, we manually validated the question type of each sample and discarded those with the wrong types. We ended up with 517 sampled questions, as shown in Table \ref{tab:questiondist}. Additionally, we randomly sampled another set of 2000 questions from the SO data dump for linguistic analysis. Since all collected questions are originally in HTML format, we removed HTML tags and stored them as plain text with their metadata (e.g., tags, view count, types, etc.) in CSV files.

\tabcolsep=0.05cm
\begin{table*}
  % \resizebox{0.50\textwidth}{!}{
  \small
  \begin{tabular}{|l||l|l|c|}
    \hline
    Properties of SO questions &  Sub-Category & Selection Criteria & \# of Questions\\\hline \hline
    % \midrule
    Type &  Conceptual & \ ( Initial question posts are divided into these three & 175  \\  & How-to & \  sub-categories by implementing an SVM classifier & 170 \\  &Debugging  & \   and manually validated afterward. ) & 172 \\
    \hline
    Popularity & Popular & Highest 10\% View Count (Avg. 28750.5) & 179 \\  & Average Popular & Average View Count (Avg. 905.3) & 165
    \\  & Unpopular & Lowest 10\% View Count (Avg. 42.1 ) & 173 \\
    \hline
     Recency  & Old & Before November 30, 2022 (the release of ChatGPT) & 266 \\ & New & After November 30, 2022 & 251 \\              
    \hline
  \end{tabular}
  % }
  \caption{Different properties of SO questions analyzed in the Manual Analysis, sub-categories for each property, and selection criteria for each sub-category of SO questions posts.}
  \label{tab:questiondist}
  \vspace{-20pt}
\end{table*}
% \vspace{-20pt}

\subsubsection{\textbf{ChatGPT Answer Collection}}

For each of the 517 SO questions, the first two authors manually used the SO question's title, body, and tags to form one question prompt\footnote{Example prompts are included in the Supplementary Material.}
and fed that to the free version of ChatGPT, which is based on GPT-3.5. We chose the free version of ChatGPT because it captures the majority of the target population of this work. Since the target population of this research is not only industry developers but also programmers of all
levels, including students and freelancers around the world, the free version of ChatGPT has significantly more users than the paid version, which costs a monthly rate of 20 US dollars. The answers generated by ChatGPT are stored in CSV files. \revision{Since ChatGPT stores the history of previous input and output of a session, a new chat session was started before feeding each question prompt to ChatGPT.}
For the additional 2000 SO questions, we developed an automated script to prompt ChatGPT with the gpt-3.5-turbo API. For each question, this script automatically extracted and concatenated its title, body, and tags based on the prompt template and stored ChatGPT answers in CSV files. \revision{Each new prompt was conducted via a new API call, which cleared the context history of previous prompts.}

% The entire data collection process is shown in Figure \ref{fig:overview}. 

%% file: sections/Opencoding.tex
\subsection{Manual Analysis}
\label{sec:opencode}
In this section, we describe the manual analysis procedure for the 517 ChatGPT answers.
% and fine-grain manual analysis.
% \todo{open coding is fine-grained manual analysis. Do you mean thematic analysis here?}  

\subsubsection{\textbf{Open Coding Procedure}}
\label{opencode}
To assess the quality and correctness of ChatGPT answers (RQ1), we used a standard NLP data labeling process~\cite{rodrigues2014sequence,xu2020review} to label the ChatGPT answers at the sentence level. Over the course of five weeks, the first three authors met six times to generate, refine, and finalize the codebooks to annotate the ChatGPT answers. 
First, the first two authors familiarized themselves with the data. Each author independently labeled five ChatGPT answers at the sentence level and took notes about their observations. The two authors met to review their labeling notes and performed thematic analysis~\cite{braun2006using,guest2011applied} to categorize the labels into four themes---{\em Correctness}, {\em Consistency}, {\em Comprehensiveness}, {\em Conciseness}. Then, they developed the initial codebook, relabeled the previous five ChatGPT answers based on the codebook, and met the other co-authors to resolve the disagreements and refine the codebook.  %Authors 1 and 2 labeled the previous 5 ChatGPT answers line-by-line and developed the initial codebook with common codes they observed in the 5 categories. 
After this step, the codebook contained 24 codes in the four themes. 

The first two authors then moved on and labeled 20 new ChatGPT answers independently based on the codebook. Since one text span in an answer may suffer from multiple quality issues, the labeling is essentially a multi-label, multi-class classification where labels are not mutually exclusive. Therefore, we cannot use Cohen's Kappa to measure the agreement level between labelers. Instead, we used Fleiss's Kappa~\cite{fleiss1971measuring} score. The initial score was 0.45, which was not high enough to proceed to label more answers. Thus, the authors met again to discuss the labeling. They carefully reviewed each label in the answers and resolved the conflicts. They further refined the codebook by merging redundant codes, improving the definitions of ambiguous codes, and introducing new codes. 
%and compared them with the codebook to decide (1) if the labelings are correct, (2) if there is a better code to describe a group of labelings, and (3) if there is an opportunity to merge redundant codes.  
% and Krippendorff's Alpha is 0.22. 
%After discussion, the codebook was extended to 31 codes in 5 categories. 
%Together, they reviewed 5 answers labeled by each of them. 
%Using these three criteria, they identified the conflicts. 
At the end of this step, there were 21 codes in the codebook. 

With the refined codebook, the first two authors re-labeled 10 of the previous 20 answers and confirmed the agreement. %They regrouped to review the labelings and identify new conflicts. At this meeting, 
 Except for disagreement about the definition and usage of 2 codes in \textit{correctness} category, no new disagreement was discovered. At this point, Fleiss's Kappa score was 0.79. Next, the first two authors met the co-authors to review and refine the current codebook and labelings. 
%A new conflict on using a separate codebook for code answers arise in this meeting. The three authors disputed that by reviewing ChatGPT answers with code and decided to use a separate codebook for annotating the coding parts of ChatGPT answers.
After this meeting, the codebook was refined to 19 codes.  

Finally, the first two authors labeled 20 new ChatGPT answers with the refined cookbook and arrived at a Fleiss's Kappa score of 0.83, which implies substantial agreement. 
% and Krippendorff's Alpha is 0.80.
With this codebook, the first two authors split the remaining ChatGPT answers and labeled them separately.  The whole labeling process took about 216 person-hours.

\subsubsection{\textbf{Definitions and Discussion of Codebook}}
\label{sec:def_codebook}

The codebook developed in the previous section contains a wide range of fine-grained codes that are used to develop a taxonomy of issues in ChatGPT answers (RQ2). We give a quick overview of these codes below. Section \ref{result:manual} provides more details.

For \textbf{\textit{Correctness}},  we compared ChatGPT answers with the accepted SO answers and also resorted to other online resources such as blog posts, tutorials, and official documentation. Our codebook includes four types of correctness issues--- \textit{Factual, Conceptual, Code}, and \textit{Terminological} errors. 
% A sentence is considered \textit{Factually Incorrect} when ChatGPT states some fabricated or untruthful information about existing knowledge. For example, claiming a certain API solves a problem when it does not, fabricating non-existent links, stating untruthful explanations, etc. On the other hand, it is considered \textit{Conceptually} incorrect if ChatGPT fails to understand the underlying context and concept of the question. For example, if a user asks how to use public and private access modifiers and ChatGPT talks about the benefits of encapsulation in C++, it is considered as \textit{Conceptually Incorrect}. And lastly, for a sentence to be \textit{Terminologically Incorrect}, it has to showcase wrong usages of correct terminology or any use of incorrect terminology, e.g., talking about \textit{perl} in a sentence where the question is a JavaScript question, using fabricated terminology, etc. 
Specifically, for incorrect code examples embedded in ChatGPT answers, we identified four types of code errors---\textit{Syntax} errors, \textit{Wrong Logic}, \textit{Wrong API/Library/Function Usage}, and \textit{Incomplete Code}. \revision{An answer is considered fully correct if it does not contain any of these errors, i.e., \textit{Factual, Conceptual, Code}, or \textit{Terminological} errors.} 

For \textbf{\textit{Consistency}}, we measured the consistency between ChatGPT answers and the accepted human-written answers on Stack Overflow. Note that inconsistency does not imply incorrectness. A ChatGPT answer can be different from an accepted human answer, but it can still be correct. Five types of inconsistencies emerged from the manual analysis---\textit{Factual Inconsistency, Conceptual Inconsistency, Terminological Inconsistency, Coding Inconsistency}, and \textit{Different Number of Solutions} (e.g., ChatGPT provides four solutions where SO gives only one).

For \textbf{\textit{Conciseness}}, three types of conciseness issues were identified and included in the codebook---\textit{Redundant}, \textit{Irrelevant}, and \textit{Excess} information. \textit{Redundant} sentences reiterate information stated in the question or in other parts of the answer. \textit{Irrelevant} sentences talk about concepts that are out of the scope of the question being asked. And lastly, \textit{Excess} sentences provide information that is not required to understand the answer.

% For example, in a question about how to fix the ending condition in a while loop, if the answer gives an elaborate description of the purpose of loops in programming, that is considered as \textit{Excess}. 

%Unlike the three characteristics above, \textbf{\textit{Comprehensiveness}} is labeled at the post level, rather than the sentence level. 
\textbf{\textit{Comprehensiveness}} is an overall assessment of the entire answer. Thus, the codebook only includes two codes---\textit{Comprehensive}, and \textit{Not Comprehensive}. To consider an answer to be comprehensive, it needs to fulfill two requirements--(1) all parts of the question are addressed in the answer, and (2) a complete solution is provided in the answer. 

%% file: sections/Linguistic.tex
\subsection{Linguistic Analysis }
\label{sec:linguistic}

Previous studies show that user preference and acceptance of an SO answer
can depend on the underlying  
emotion, tone, linguistic style, and sentiment in the answer~\cite{shneiderman1980software,calefato2015mining,bazelli2013personality}. In this section, we describe the automated methods utilized to determine linguistic features and sentiments of ChatGPT answers. 

\subsubsection{\textbf{Linguistic Characteristics}}

% \todo{Follow the LIWC structure from the ``SyntheticLies'' paper, and "On the Personality Traits of StackOverflow Users" paper, and Bogdan's LIWC description from "A large-scale, in-depth analysis of developers’ personalities in the Apache ecosystem"}

We employed a widely used tool called  Linguistic Inquiry and Word Count (LIWC)~\cite{pennebaker2015development} %pronounced \textit{luke})
to analyze the linguistic features of ChatGPT and SO answers. LIWC %\footnote{https://www.liwc.app/static/documents/LIWC2015\%20Manual\%20-\%20Development\%20and\%20Psychometrics.pdf} 
is a psycholinguistic database that provides a dictionary of validated psycholinguistic lexicons in pre-determined categories that are psychologically meaningful. LIWC counts word occurrence frequencies in each category that holds important information about the emotional, cognitive, and structural components associated with text or speech. 
LIWC has been used to study AI-generated misinformation~\cite{zhou2023synthetic}, emotional expressions in social media posts~\cite{kramer2012spread}, the success of human answers~\cite{bazelli2013personality}, etc. In our work, we considered the following categories:

% \hspace{-20pt}
\begin{itemize}
    \item \textbf{Linguistic Styles:} We considered four attributes related to linguistic styles---\textit{Analytical Thinking} (complex thinking, abstract thinking), \textit{Clout} (power, confidence, or influential expression), \textit{Authentic} (spontaneity of language), and \textit{Emotional Tone}. 

    \item \textbf{Affective Attributes:} Affective attributes capture expressions and features related to emotional status. They include \textit{Affect} (overall emotional expressions, e.g., ``happy'', ``cried''), \textit{Positive Emotion} (e.g., ``happy'', ``nice''), and \textit{Negative Emotion} (e.g., ``hurt'', ``cried'').
    
    \item \textbf{Cognitive Processes:} Cognitive processes represent features that are related to cognitive thinking and processing, e.g., causation, knowledge, insight, etc. For this category, we considered \textit{Insight} (e.g., ``think'', ``know''), \textit{Causation} (e.g., ``because''), \textit{Discrepancy} (e.g., ``should'', ``would''), \textit{Tentative} (e.g., ``perhaps''), \textit{Certainty} (e.g., ``always''), and \textit{Differentiation} (e.g., ``but'', ``else'').

    \item \textbf{Drives Attributes:} Drives capture expressions that show the need, desire, and effort to achieve something. For this category, we considered \textit{Drives}, \textit{Affiliation} (e.g., ``ally'', ``friend''), \textit{Achievement} (e.g., ``win'', ``sucess''), \textit{Power} (e.g., ``superior''), \textit{Reward} (e.g., ``prize'', ``benefit''), and \textit{Risk} (e.g., ``danger'', ``doubt'').  

    \item \textbf{Perceptual Attributes:} This category captures the attributes that are related to \textit{Perceive, See, Feel}, or \textit{Hear}. 

    \item \textbf{Informal Attributes:} This category captures the causality in everyday conversations. The attributes in this category include \textit{Informal Language, Swear Words, Netspeak} (e.g., ``btw, lol''), \textit{Assent} (e.g., ``OK'', ``Yeah''), \textit{Nonfluencies} (e.g., ``er'', ``hmm''), and \textit{Fillers} (e.g., ``I mean'', ``you know'').
\end{itemize}

We used LIWC to compute word frequency in each of the categories for 2000 ChatGPT answers and the corresponding human answers from Stack Overflow. For ease of understanding, we computed the relative differences (\textit{RD}) in linguistic features between 2000 pairs of ChatGPT and SO answers from the computed average word frequencies in each category. 
\vspace{-2pt}
% \[ RD= \frac{ChatGPT\ \ avg.\ \ frequency - SO\ \ avg.\ \ frequency}{SO\ \ avg.\ \ frequency}\]
% \[ RD= \frac{CF - SF}{SF}\]

\begin{equation*}
    RD = \frac{ChatGPT\ \ avg.\ \ frequency - SO\ \ avg.\ \ frequency}{SO\ \ avg.\ \ frequency}
\end{equation*}

% \noindent where CF refers to ...., and SF refers to ....

\subsubsection{\textbf{Sentiment Analysis}}
Lexicon-based LIWC evaluates linguistic characteristics based on psycholinguistic features and captures the sentiment of texts only based on overall polarity. Hence, LIWC is insufficient when it comes to capturing the intensity of the polarity~\cite{sentimentmedium}. Moreover, LIWC can not capture sarcasm, irony, misspelling, or negation, which is necessary to analyze sentiment in human-written texts on Q\&A platforms. Therefore, we employed a machine learning algorithm to further evaluate and compare the underlying sentiment portrayed in the ChatGPT answers and human answers. Specifically, we used a RoBERTa-based sentiment analysis model from Hugging Face~\cite{huggingface}. % which is fine-tuned on software engineering texts\footnote{Cloudy1225/stackoverflow-roberta-base-sentiment}. 
This model is pre-trained on a Twitter corpus %and the ``twitter-roberta-base-sentiment'' model\footnote{cardiffnlp/twitter-roberta-base-sentiment} 
and is then finetuned with the 4423 annotated SO posts from Calefato et al.~\cite{calefato2018sentiment}. This well-balanced dataset has 35\% posts with positive sentiment, 27\% of posts with negative sentiment, and 38\% of posts with neutral sentiment. 

% This model predicted the sentiment (positive, neutral, or negative) of each of the 2000 ChatGPT and SO answers. 

% Answers with token size higher than 512 were split into overlapping segments and predictions on all segments were merged with a voting mechanism to accumulate the final predictions for those answers.

%% file: sections/userstudy.tex
% \begin{tcolorbox}[breakable,notitle,boxrule=0.5pt,colback=yellow!15,colframe=yellow!105]

%     \textbf{RQ7.} Can users differentiate between machine-generated and human answers?
%     \newline

%     \textbf{RQ8.} Can users identify misinformation in machine-generated answers?
%     \newline
    
%     \textbf{RQ7.} Do users prefer ChatGPT over SO \todo{solely based on answer quality/without knowing the source of the answer}?
%     \newline

%     \textbf{RQ10.} Do users need additional support with ChatGPT answers compared to SO?
    
% \end{tcolorbox}

\subsection{User Study}
\label{sec:userstudy}

To understand programmers' perception of ChatGPT answers and human answers, we conducted a within-subjects user study with 12 participants. Our goal is to observe how programmers assess those answers and which kind of answers they prefer. %In the study, the participants saw a series of Stack Overflow questions with two sets of answers and rated their characteristics. 
% along with stating their preference between the answers. 
% We also asked users to identify the incorrect answer, guess the machine-generated answer, and state their preference for one answer over the other. In the semi-structured interview, we discussed the \textit{Why} of their preferences and the \textit{How} of their heuristics for identification.

\subsubsection{\textbf{Participants}}
For the user study, we recruited 12 participants (3 female, 9 male) with programming backgrounds. 7 participants were graduate students, 4 participants were undergraduate students, and 1 participant was a software engineer from the industry. The participants were recruited by word of mouth.
Participants rated their programming expertise by answering multiple-choice questions with five options---{\em Novice, Beginner, Competent, Proficient, and Expert}. Eight participants self-reported as proficient, three as competent, and one as beginner. 
\revision{Since some participants may be modest about their programming skills, we also collected the number of years of programming experience. Four participants had three years of experience, one had four years, one had five years, two had six years, one had seven years, and three had eight years of programming experience.} %However, the self-reported expertise did not match the programming years since programmers with fewer years of experience can become more efficient programmers depending on the level of practice. Hence we collected both of these metric.}
%We also asked participants about their expertise of using ChatGPT. Three participants self-reported as proficient with ChatGPT, six were competent, two were novices, and one was a beginner
Additionally, we asked participants how often they use ChatGPT and how often they use SO.
%They answered answering multiple-choice questions with 5 options--- Never, Seldom, Some of the Time, Very Often, and All the Time.
For ChatGPT, three answered \textit{very often}, three answered \textit{some of the time}, two answered \textit{seldom}, and four answered \textit{never}. 
For SO, four participants answered \textit{all the time}, five answered \textit{very often}, two answered \textit{some of the time}, and one answered \textit{seldom}. 
%Participation was voluntary and there was no compensation for participating in the study.

\subsubsection{\textbf{SO Question Selection}} We randomly sampled eight questions from our manual analysis dataset. ChatGPT gave incorrect answers to five questions and correct answers to three questions. One question was about C++, one about PHP, two about HTML/CSS, three about JavaScript, and one about Python. 

\subsubsection{\textbf{Protocol}} In this user study, we asked participants to complete a sequence of decision-making tasks to verify and assess the quality of machine and human-generated answers to programming questions. The tasks were designed to capture user perception and preference for human and machine-generated answers. In each task, we asked the participants to verify and assess a ChatGPT answer and a human answer to a SO question and rate the correctness and quality of each answer. Moreover, for each task, the participants were asked to mark which answer they preferred and guess which answer was generated by ChatGPT. The step-by-step procedure for each task is described below.  

Each user study started with consent collection and an introduction to the study procedure. Then, the participants started the study tasks by reading each SO question and rating their familiarity with the topic asked in the question (5-point Likert Scale\revision{~\cite{norman2010likert}}).
\revision{Familiarity with a specific programming topic is not directly related to years or hours of programming experience, as programmers tend to be more familiar with topics they have used more recently. Therefore, we resorted to the self-reporting method.}
Then, they were presented with an answer to the question. This answer is either generated by ChatGPT or written by a human programmer on Stack Overflow.  Then, they were asked to answer a series of 5-point scale survey questions to assess the correctness, comprehensiveness, conciseness, and usefulness of this answer. Then, they were presented with the other answer and asked to answer the same set of survey questions. Then, they were asked to select which answer they prefer, which answer they believe is generated by ChatGPT, and how confident they are about their choices. We repeated this process for all eight SO questions and randomized the order of ChatGPT and human answers for each question. For ease of running this study, all SO questions, answers, survey questions, and instructions were encoded into a Qualtrics survey.\footnote{The survey is included in Supplementary Material} 
%Each SO question (each task) had its individual page in the survey. The sequence of survey questions 
% for each task is in the following order---\textbf{(1)} participant's expertise in the topic (e.g., C++, PHP, etc.) of the SO question (5-point Likert Scale),
% \textbf{(2)} the SO question, 
% \textbf{(3)} an answer generated by ChatGPT or an accepted SO answer written by human\footnote{The order of the two answers is randomized. If ChatGPT generates the first answer, the second answer is from Stack Overflow, and vice versa.}, 
% \textbf{(4)} 5-point scale questions for assessing the correctness, comprehensiveness, conciseness, and usefulness of the first answer, 
% \textbf{(5)} the other answer,   
% \textbf{(6)} 5-point scale questions for assessing the correctness, comprehensiveness, conciseness, and usefulness of the second answer, 
% \textbf{(7)} A multiple choice question to capture user preference (choices--- Answer 1, Answer 2), 
% \textbf{(8)} A multiple choice question to capture user guess of machine-generated answer (choices--- Answer 1, Answer 2),  
% \textbf{(9)} A multiple choice question to capture user perception of incorrect answer (choices--- Answer 1, Answer 2, none of the above),  
% \textbf{(10)} Confidence rating (5-point Likert Scale).  

The human answers and ChatGPT answers were presented with the same text format and style (e.g., font type, font size, code format, etc.), so participants could not easily tell them apart just based on formatting and visual styles. 
%All participants observed the SO questions in the same order. However, 
Participants were allowed to skip to the next SO question if they were not familiar with the topic of a certain SO question. The order of ChatGPT and human answers was assigned randomly (i.e., not always Answer 1 was ChatGPT answer). Additionally, participants were encouraged to refer to external resources, such as Google search, tutorials, and API documentation, to verify the correctness of the given answers. In the verification process, to prevent participants from running into the same human answer on Stack Overflow or getting the same answer from ChatGPT, the participants were not allowed to search on Stack Overflow, open a Stack Overflow page returned by Google Search, or ask the same question to ChatGPT.
\revision{Apart from accessing ChatGPT and SO for the same question, participants were allowed to validate the code generated by ChatGPT in any local IDE, online sandbox, or online code editor at their convenience.}
Each participant was given 20 minutes to examine and rate answers to SO questions. Participants were made aware that finishing all eight questions was not required and were encouraged to aim for comprehensiveness and quality instead of the number of examined answers. 
All participants used up the given 20 minutes of time in the study. On average, participants assessed the correctness and quality of the answers to 5 questions. 

\subsubsection{\textbf{Semi-Structured Interview}}
The survey was followed by a lightweight semi-structured interview. Each interview took about 10 minutes on average.
During the interview, we reviewed the participant's responses to the survey together with the participant and asked them why they preferred one answer over the other. 
%After this, we reviewed the question where participants guessed which answer was machine-generated. 
Then, we asked the participants about their heuristics to identify the ChatGPT answer before revealing the correct answer to them. If the participants were correct, we asked a follow-up question about the characteristics of ChatGPT answers that influenced their decision. Lastly, we asked how they determined the incorrect information in an answer. We also asked follow-up questions such as why they failed to identify some misinformation, what the main challenges were in verifying the correctness, what additional tool  support they wish to have, etc.\footnote{A complete list of interview questions are included in the Supplementary Material.}
% In the end, after reviewing all answered questions, we asked for their general view and perception pros and cons of machine-generated answers in SE tasks. 

\subsubsection{\textbf{Qualitative Analysis of the Interview Transcripts}}
% We conducted a quantitative analysis of the survey results to calculate the percentage of time participants prefer SO answers over ChatGPT or vice versa, how successfully users identify incorrect answers, and machine-generated answers, and analyzed their overall ratings for the answers. 
%To understand what factors play roles in user preference, and the heuristics of the identification of machine-generated and incorrect answers, 
The first author transcribed the audio recordings and labeled all 12 transcripts following the open coding method~\cite{hancock2001introduction}. The author labeled all insightful responses that mentioned factors related to participants' preferences, the heuristics used by the participants, the obstacles they faced, and the tool support they wished to have. After this step, the author did a thematic analysis~\cite{braun2006using,guest2011applied} to group the low-level labels into high-level patterns and themes. The final codebook for thematic analysis contains 5 themes and 21 patterns.\footnote {We have included the codebook in Supplementary Material.} The overall process took about six person-hours.

% \todo{
% \begin{itemize}
%     \item How do the users differentiate between human and machine-generated answers? Do they look for specific language styles (e.g., tone, sentiment, organization) or answer quality (e.g. comprehensiveness, length, etc. )?

%     \item If a user could not identify the wrong answer/misinformation, then what are the reasons for that? What are the properties of the machine-generated answers that mimicked accurate information and prompted their actions?

%     \item If a user could not identify the wrong answer/misinformation, then what additional feature do they think might have helped them to distinguish wrong information from correct information?

%     \item If a user successfully found all/most of the incorrect answers, then what are the features/signs of the machine-generated answer that helped them do that? What were their tactics? 

% \end{itemize}
% }

%% file: sections/Results_manual.tex
\section{Manual Analysis Results}
\label{result:manual}
This section presents the results and findings for RQ1-RQ3. 

% We sorted the percentage of incorrectness \todo{I get confused here. Why calculating percentage? Isn't it a binary question?} \new{referring back 3.2.3, the percentage of incorrectness in each answer can be anything from 0 to 100\%. We pick the absolute threshold to make the correctness status binary. e.g., below 0.5\% incorrect ,above 0.5\% correct answer. } \todo{why not simply consider the number of incorrect information labeled in an answer}among all answers and found that the lowest percentage of incorrectness is 0.5\% and used it as our threshold for determining the correct or incorrect status of an answer. We calculated the threshold of inconsistency to be 0.9\% in a similar manner and all answers with inconsistencies less than this threshold are considered consistent.  

\subsection{RQ1: Overall Correctness and Quality}
%To evaluate the overall correctness, we computed the number of correct and incorrect labels for 517 answers.
% \todo{I find the original writing is too verbose, so I rewrote it. I find it unnecessary to spell out the frequency of each label here and report the results according to the labels. Just want you to know this. You can delete this comment after reading it.}
Our results show that, among the 517 ChatGPT answers we labeled, 52\% of them contain incorrect information, 78\% are inconsistent from human answers, 35\% lack comprehensiveness, and 77\% contain redundant, irrelevant, or unnecessary information. 
Moreover, on average, ChatGPT answers and human answers contain 266.43 tokens (\(\sigma\)=87.99) and 213.80 tokens (\(\sigma\)=246.04) respectively. The mean difference of 52.63 tokens is statistically significant (paired t-test: p-value<0.001). Table \ref{tab:manual_all} shows our manual analysis results. %\todo{This result is a bit short. Try to add some elaborations.} \new{I don't think there is anything more to add here since all quantitative results for overall correctness/incorrectness are included already. The next 2 RQs cover the other aspects. Since this RQ focuses on overall result, it's supposed to be small. At best I can add definitions and examples for comprehensiveness }

% \todo{remove this paragraph}Figure \ref{fig:useful} shows the usefulness of 517 answers as rated by annotators. The result shows that 38\% of the answers are very \textbf{\textit{Useful}} and 15\% of the answers are moderately \textbf{\textit{Useful}}. 

% In summary, the overall correctness and quality of ChatGPT answers differ from SO answers. 
\vspace{-4pt}
\begin{eqbox}[\textbf{Finding 1} ]
 More than half of ChatGPT answers contain incorrect information, 77\% of ChatGPT answers are verbose, and 78\% of ChatGPT answers contain inconsistencies with human answers. However, ChatGPT answers are comprehensive and cover different aspects of the questions and answers. 
\end{eqbox}
% About 50\% of ChatGPT answers are incorrect, verbose, and less useful, but comprehensive.

% \tabcolsep=0.05cm
\begin{table*}[]
    \centering
    \small
    % \resizebox{0.50\textwidth}{!}{
\begin{tabular}{|c|c||c|c|c|c|} \hline

%%%%%% Title2 row starts here
&  & Correct  & Consistent  & Comprehensive  & Concise  \\ \hline
&  & Yes \hspace{0.2cm} No & Yes \hspace{0.2cm} No & Yes \hspace{0.2cm} No & Yes \hspace{0.2cm} No \\ \hline\hline
%%%%%% Row Popular starts here
Popularity &
\begin{tabular}{c}
    Popular \\ Avg. Popular \\ Not Popular
\end{tabular} & \cellcolor{cyan!15}
\begin{tabular}{c} 0.55 \hspace{0.2cm} 0.45 \\ 0.46 \hspace{0.2cm} 0.54 \\ 0.42 \hspace{0.2cm} \underline{0.58} \\
\end{tabular} &
\begin{tabular}{c} 0.21 \hspace{0.2cm} 0.79 \\ 0.22 \hspace{0.2cm} 0.78 \\ 0.25 \hspace{0.2cm} 0.75 \\
\end{tabular} &
\begin{tabular}{c} 0.64 \hspace{0.2cm} 0.36 \\ 0.64 \hspace{0.2cm} 0.36 \\ 0.66 \hspace{0.2cm} 0.34 \\
\end{tabular} & \cellcolor{cyan!15}
\begin{tabular}{c} 0.16 \hspace{0.2cm} \underline{0.84} \\ 0.26 \hspace{0.2cm} 0.74 \\ 0.28 \hspace{0.2cm} 0.72 \\
\end{tabular} \\ \hline
%%%%%% Row Type starts here
Type &
\begin{tabular}{c}
    Debugging \\ How-to \\ Conceptual
\end{tabular} &
\begin{tabular}{c} 0.45 \hspace{0.2cm} \underline{0.55} \\ 0.47 \hspace{0.2cm} 0.53 \\ 0.48 \hspace{0.2cm} 0.52 \\
\end{tabular} & \cellcolor{cyan!15}
\begin{tabular}{c} 0.17 \hspace{0.2cm} \underline{0.83} \\ 0.21 \hspace{0.2cm} 0.79 \\ 0.28 \hspace{0.2cm} 0.72 \\
\end{tabular} &
\begin{tabular}{c} 0.63 \hspace{0.2cm} 0.37 \\ 0.67 \hspace{0.2cm} 0.33 \\ 0.64 \hspace{0.2cm} 0.36 \\
\end{tabular} & \cellcolor{cyan!15}
\begin{tabular}{c} 0.40 \hspace{0.2cm} 0.60 \\ 0.13 \hspace{0.2cm} \underline{0.87} \\ 0.16 \hspace{0.2cm} \underline{0.84} \\
\end{tabular} \\ \hline

%%%%%% Row Time starts here
Recency &
\begin{tabular}{c}
    Old \\ New 
\end{tabular} & \cellcolor{cyan!15}
\begin{tabular}{c} 0.53 \hspace{0.2cm} 0.47 \\ 0.42 \hspace{0.2cm} \underline{0.58} \\ 
\end{tabular} &
\begin{tabular}{c} 0.22 \hspace{0.2cm} 0.78 \\ 0.22 \hspace{0.2cm} 0.78 \\ 
\end{tabular} &
\begin{tabular}{c} 0.68 \hspace{0.2cm} 0.32 \\ 0.61 \hspace{0.2cm} 0.39 \\ 
\end{tabular} & \cellcolor{cyan!15}
\begin{tabular}{c} 0.17 \hspace{0.2cm} \underline{0.83} \\ 0.29 \hspace{0.2cm} 0.71 \\ 
\end{tabular} \\ \hline

%%%%%% Row Overall starts here
Overall & \_ &
\begin{tabular}{c} 0.48 \hspace{0.2cm} 0.52 \\
\end{tabular} &
\begin{tabular}{c} 0.22 \hspace{0.2cm} 0.78 \\
\end{tabular} &
\begin{tabular}{c} 0.65 \hspace{0.2cm} 0.35 \\ 
\end{tabular} &
\begin{tabular}{c} 0.23 \hspace{0.2cm} 0.77 \\
\end{tabular} \\ \hline

\end{tabular}
% }
    % \vspace{-20pt}
    \caption{Percentage distribution of ChatGPT answers for all 4 correctness and quality issues (Correctness, Consistency, Comprehensiveness, and Conciseness) across 3 properties of question posts (Popularity, Type, and Time). The statistically significant (Pearson's Chi-square Test: p-value<0.05) relations are highlighted in blue.}
    \label{tab:manual_all}
    \vspace{-20pt}
\end{table*}

% \begin{figure}[]
%     \centering
%     \includegraphics[width=0.30\textwidth]{figures/overall_usefulness_2.eps}
%     \vspace{-15pt}
%     \caption{Usefulness of ChatGPT answers rated by annotators}
%     \label{fig:useful}
%     \vspace{-10pt}
% \end{figure}

\vspace{-5pt}
\subsection{RQ2: A Taxonomy of Fine-Grained Issues in ChatGPT Answers}
%We compiled the percentage of answers based on finer issues identified from the manual labeling of the answers. 
Our thematic analysis reveals four types of \textit{\textbf{incorrectness}} in ChatGPT answers---\textit{Conceptual} (54\%), \textit{Factual} (36\%), \textit{Code} (28\%) and \textit{Terminology} (12\%) errors. Note that these errors are not mutually exclusive. Some answers have more than one of these errors. 
\textit{Factual} errors occur when ChatGPT states some fabricated or untruthful information about existing knowledge, e.g., claiming a certain API solves a problem when it does not, fabricating non-existent links, untruthful explanations, etc. On the other hand, \textit{Conceptual} errors occur if ChatGPT fails to understand the question. For example, the user asked how to use public and private access modifiers, and ChatGPT answered the benefits of encapsulation in C++. \textit{Code} errors occur when the code example in the answer does not work, or cannot provide a desired output. And lastly, \textit{Terminology} errors are related to wrong usages of correct terminology or any use of incorrect terminology, e.g., \textit{perl} as a header of \textit{Python} code.

% This result suggests that more than half of the errors are made due to ChatGPT not understanding the context of the questions. When labeling the answers, we noticed ChatGPT either can not understand the question. In cases when it can understand the question, it fails to show an understanding of how to solve the problem. This contributes to the high number of \textit{Conceptual} errors. The rate of \textit{Factual} error is lower than \textit{Conceptual} error but still composes a large portion of incorrectness. Observation while labeling showed that the factual errors that ChatGPT makes are mostly due to providing the wrong links or non-existent links, fabricating API or documentation, stating that an API/library/function can achieve something while their functionalities in the real world are completely different, etc. The third type of error was very small compared to the other two errors. \textit{Terminology} error mainly occurs when ChatGPT uses terminology from one language or concept in another. Most of the \textit{Terminology} we observed are mainly using the wrong language name on top of code solutions. e.g., using ``Perl'' as a language name in a header of a ``C++'' code. 

Specifically, for code errors, our analysis reveals four types of code errors---wrong logic (48\%), wrong API/library/function usage (39\%), incomplete code (11\%), and wrong syntax (2\%). Again, some generated code has more than one of these errors. 
 Logical errors are made by ChatGPT when it can not understand the problem, fails to pinpoint the exact part of the problem, or provides a solution that does not solve the problem. For example, in many debugging instances, we found that ChatGPT tried to resolve one part of the given code, whereas the problem lied in another part of the code. One such example is provided in Appendix A. We also observed that ChatGPT often fabricated APIs or claimed certain functionalities that were wrong.

% \vspace{-6pt}
 \begin{eqbox}[\textbf{Finding 2}]
Many answers are incorrect due to ChatGPT's incapability to understand the underlying context of the question being asked. Yet ChatGPT makes fewer factual errors compared to conceptual errors. 
\end{eqbox}

\vspace{-10pt}

\begin{eqbox}[\textbf{Finding 3}]
ChatGPT rarely makes syntax errors for code answers. The majority of the code errors are due to applying wrong logic or implementing non-existing or wrong API, library, or functions.  
\end{eqbox}

\revision{The ChatGPT answers that have no statements annotated as factual, conceptual, code, or terminological errors, are considered to be correct. We found that 48\% of the ChatGPT answers had an absence of any type of fine-grained errors identified in the manual analysis.}

 % Our result suggests that most of the codes are incorrect due to logical errors. From our observation, this is analogous to conceptual error for the non-code part of the answers. Most of the logical errors are made by ChatGPT due to the fact that it can not understand the problem as a whole, fails to pinpoint the exact part of the problem, or provides a solution that does not solve the problem. For example, in many debugging instances, we found that ChatGPT tries to resolve one part of the given code, whereas the problem lies in another part of the code. Also, in many cases, ChatGPT makes discernible logical errors that no human expert can make. For example, setting a loop ending condition to be equal to something that is never true or never false (e.g. \(while(i<0  and i>10)\)). The next major reason for code error is the wrong usage of API/library/function. This is again analogous to factual error of non-code parts of the answer. 
 % This error mainly occurs when ChatGPT fabricates API/library/function, or claims certain functionalities that are wrong, using API/library/functions to solve a problem that requires different sets of API/library/functions, etc.
 % We found a small (11\%) percentage of codes that are incomplete and hence do not solve the question at hand. A very small percentage (2\%) of codes have a syntax error. While our observation reveals that ChatGPT makes \textit{Terminology} errors sometimes, we rarely found syntax errors in the code answers. 

Among the answers that are \textit{\textbf{Not Concise}}, 46\% of them have {\em Redundant} information, 33\% have {\em Excess} information, and 22\% have {\em Irrelevant} information. For {\em Redundant} information, during our labeling process, we observed that many of the ChatGPT answers repeat the same information that is either stated in the question or stated in other parts of the answers. 
% This is the most common behavior among all parts that we marked as not concise. 
For {\em Excess} information, we observed a handful of cases where ChatGPT unnecessarily gives background information such as long definitions, or writes something at the end of the answer that does not add any necessary information to understand the solution. 
Lastly, many answers contain {\em Irrelevant} information that is out of context or scope of the question. In answers with conceptual errors, we observed this behavior more often. There are answers that have a combination of more than one of these conciseness issues. An example of a verbose ChatGPT response is provided in Appendix B. 

And lastly, for inconsistency with human answers, we found five types of \textit{\textbf{Inconsistencies}}---\textit{Conceptual} (67\%), \textit{Factual} (44\%), \textit{Code} (55\%), \textit{Terminology} (6\%), and \textit{Number of Solutions} (42\%). 
The first four types of inconsistencies occur for the same reason as incorrectness. The only difference is that inconsistency does not always mean incorrectness, as explained in Section \ref{sec:def_codebook}. Similar to incorrectness, conceptual inconsistencies are higher than factual inconsistencies. %However, the code generated by ChatGPT are significantly more inconsistent compared to incorrect code. 
Our observation also reveals that ChatGPT-generated code is very different from human-written code in format, semantics, syntax, and logic. This contributes to the higher number of {\em Code} inconsistencies. 
The \textit{Number of solutions} inconsistency is very prominent as ChatGPT often provides many additional solutions to solve a problem. %In a sense, inconsistency due to \textit{Number of solutions} can point towards the reason behind the prolonged answers of ChatGPT. 

\subsection{RQ3: Effects of Question Type}
To evaluate the relationship between question types and ChatGPT answer quality, we calculated the percentage of each label across all categories for each question type.
% in a similar manner described in (RQ1).  
As our data is entirely categorical, we evaluated the statistical significance of the relationship between each question type and each of the four label categories with Pearson's Chi-square test. Table \ref{tab:manual_all} highlights all relationships that are statistically significant (p-value<0.05). Our results show that \textit{\textbf{Question Popularity}} and \textit{\textbf{Recency}} have a statistically significant impact on the \textit{\textbf{Correctness}} of answers. Specifically, answers to popular questions and questions posted before November 2022 (the release date of ChatGPT) have fewer incorrect answers than answers to other questions. This implies that ChatGPT generates more correct answers when it has more information about the question topic in its training data.  
Although \textit{\textbf{Debugging}} questions have more incorrect ChatGPT answers, the difference is not statistically significant. This indicates that \textit{\textbf{Question Type}} does not affect the \textit{\textbf{Correctness}} of ChatGPT answers. 
% Whereas how popular the question topic is and how old the question is has a significant effect on the answer's correctness. 

Additionally, we found a statistically significant relationship between \textit{\textbf{Question Type}} and \textit{\textbf{Inconsistency}}. Since there are often multiple ways to debug and fix a problem, the inconsistencies between human and ChatGPT-generated answers for \textit{\textbf{Debugging}} questions are higher, with 83\% of \textit{\textbf{inconsistent}} answers.  Our observation aligns with this result too. While labeling the answers, we found that almost half of the correct \textit{\textbf{Debugging}} answers use different logic, API, or library to solve a problem that produces the same output as human answers. 

Our results also show that ChatGPT answers are consistently \textit{\textbf{Comprehensive}} for all categories of SO questions and do not vary with different \textit{\textbf{Question Type, Recency,}} or \textit{\textbf{Popularity}}.

% \begin{eqbox}[\textbf{Finding 4}]
% The Popularity, Type, and Recency of the SO questions affect the correctness and quality of ChatGPT answers. Answers to more Popular and older posts are less incorrect and more verbose. \textit{Debugging} posts are more incorrect and inconsistent, but less verbose, and answers to \textit{Conceptual} and \textit{How-to} questions are the most verbose.
% \end{eqbox}

Moreover, our analysis shows that answers to all kinds of questions, irrespective of the \textit{\textbf{Type, Recency,}} and \textit{\textbf{Popularity}}, are consistently verbose. Yet answers to different kinds of questions indeed have statistical differences in verbosity. 
Specifically, answers to \textit{\textbf{Popular}} questions are \textit{\textbf{Not Concise}} 84\% of the time, while answers for \textit{\textbf{Average}} and \textit{\textbf{Not Popular}} questions are \textit{\textbf{Not Concise}} 74\% and 72\% of the time. This suggests that for questions targeting popular topics, ChatGPT has more information on them and adds lengthy details. We found the same pattern for \textit{\textbf{Old}} questions. 
Answers to 
% \todo{Old is in italic font in this sentence but in bold in the next sentence. I have observed this kind of styling inconsistencies here and there. This looks very unprofessional to your reviewers.}
\textit{\textbf{Old}} questions (83\%) are more verbose than \textit{\textbf{New}} questions (71\%). Finally, for \textit{\textbf{Question Type}},  \textit{\textbf{Debugging}} answers are more \textit{\textbf{Concise}} (40\%) compared to  \textit{\textbf{Conceptual}} (16\%) and \textit{\textbf{How-to}} (13\%) answers, which are extremely verbose. This is because of ChatGPT's tendency to elaborate definitions for \textit{\textbf{Conceptual}} questions and to generate step-by-step descriptions for \textit{\textbf{How-to}} questions. 
% Our annotations show that for \textit{\textbf{Conceptual}} and \textit{\textbf{How-to}} answers, ChatGPT includes unnecessary details such as extra background information, irrelevant information, or multiple redundant solutions, which prolongs the answers.
% Whereas the \textit{Debugging} answers are rather to the point and often address the exact problem asked in the question without unnecessary lengthening.  

\vspace{-5pt}
\begin{eqbox}[\textbf{Finding 4}]
Popularity, Type, and Recency of programming questions affect the correctness and quality of ChatGPT answers. Answers to more \textit{Popular} and \textit{Older} posts are less incorrect and more verbose. \textit{Debugging} answers are more inconsistent but less verbose. \textit{Conceptual} and \textit{How-to} answers are the most verbose.
\end{eqbox}

% \vspace{-20pt}

%% file: sections/Result_linguistic.tex
\section{Linguistic Analysis Results}
\label{result:ling}

\begin{table}[]
    \centering
     % \footnotesize
     \small
\begin{tabular}{|l|l||l|l|} \hline
%%%%%% Title row starts here
Linguistic Features & Rel. Diff.(\%) & Linguistic Features & Rel. Diff.(\%) \\ \hline\hline
%%%%%% Row 1 starts here
\begin{tabular}{l}
    \textbf{Language Styles} \\ Analytic \\ Clout \\ Authentic \\ Tone \\
\end{tabular} &
\begin{tabular}{l} \\ 20.65*** \\ 13.01*** \\ \textcolor{orange}{-38.50}*** \\ 14.95*** \\
\end{tabular} &
\begin{tabular}{l} \textbf{Drive Attributes} \\ Drives \\ Affiliation \\ Achievement \\ Power \\ Reward \\
Risk \\
\end{tabular} & 
\begin{tabular}{l} \\ 9.53*** \\ 16.05** \\ 10.85*** \\ 22.86*** \\ 2.23 \\ \textcolor{orange} {-7.08}\\
\end{tabular} \\ \hline
%%%%%% Row 2 starts here
\begin{tabular}{l}
    \textbf{Affective Attributes} \\ Affect \\ Positive Emotion \\ Negative Emotion \\
\end{tabular} &
\begin{tabular}{l} \\ \textcolor{orange} {-6.53}** \\ 2.09 \\ \textcolor{orange}{-34.45}*** \\
\end{tabular} &
\begin{tabular}{l} \textbf{Perception Attributes} \\ Perception \\ See \\ Hear \\ Feel \\
\end{tabular} &
\begin{tabular}{l} \\ \textcolor{orange}{-26.28}*** \\ \textcolor{orange}{-34.98}*** \\ \textcolor{orange}{-16.50}* \\ 7.55 \\
\end{tabular} \\ \hline
%%%%%% Row 3 starts here
\begin{tabular}{l}
    \textbf{Cognitive Attributes} \\ Insight \\ Causation\\ Discrepancy \\ Tentative \\ Certainty \\ Differentiation\\
\end{tabular} &
\begin{tabular}{l} \\ \textcolor{orange}{-8.86}** \\ 23.94*** \\ \textcolor{orange}{-35.89}*** \\ \textcolor{orange}{-10.23}*** \\ \textcolor{orange}{-4.23} \\ \textcolor{orange}{-13.29}***\\
\end{tabular} &
\begin{tabular}{l} \textbf{Informal Attributes} \\ Informal Language\\ Swear words\\ Netspeak\\
Assent\\ Nonfluencies \\ Fillers \\
\end{tabular} & 
\begin{tabular}{l} \\ \textcolor{orange}{-53.97}*** \\ \textcolor{orange}{-71.52}** \\ \textcolor{orange}
{-60.03}***\\ \textcolor{orange}{-11.86} \\ \textcolor{orange}{-55.34}*** \\\textcolor{orange}{-82.85}***
\end{tabular} \\ \hline
\end{tabular}
    \caption{Relative Linguistic Differences (\%) between 2000 pairs of ChatGPT and human answers. Positive numbers indicate higher occurrence frequencies of linguistic features in ChatGPT answers compared to SO, and \textcolor{orange}{negative numbers} indicate lower occurrence frequencies. Numbers marked with (*) indicate differences that are statistically significant (paired t-test: *** means p-value<0.001, ** means p-value<0.01, * means p-value<0.05)  }
    \label{tab:liwc}
    \vspace{-20pt}
\end{table}
% \vspace{-40pt}

\subsection{RQ4: Linguistic Characteristics}
 Table \ref{tab:liwc} presents the relative differences in the linguistic features between ChatGPT answers and human answers. As stated in Section \ref{sec:linguistic}, relative differences capture the normalized difference in word frequencies for each linguistic feature between ChatGPT answers and human answers. Positive relative differences indicate features prominent in ChatGPT answers, and negative relative differences indicate features prominent in human answers. Our result shows several statistically significant linguistic differences between ChatGPT answers and human answers. 

First, we found that ChatGPT answers differ from human answers in terms of \textit{\textbf{language styles}}. ChatGPT answers are found to contain more words related to \textit{analytical thinking and clout expressions}. This indicates that ChatGPT answers communicate a more abstract and cognitive understanding of the answer topic, and the language style is more influential and confident. On the other hand, human answers include fewer words related to \textit{authenticity}, indicating that human answers are more spontaneous and non-regulated. 

For \textit{\textbf{affective}} attributes that capture emotional status, we found human answers contain more keywords related to emotional status. Though not statistically significant, ChatGPT answers portray more positive emotions, whereas human answers portray significantly more negative emotions than ChatGPT.
% which is the main contributor towards higher occurrence frequencies for affective attributes.

Moreover, ChatGPT answers contain significantly more \textit{\textbf{drives}} attributes compared to human answers. ChatGPT conveys stronger \textit{drives, affiliation, achievement}, and \textit{power} in its answers. We observed that many ChatGPT answers include words and phrases, such as ``of course I can help you'' and ``this will certainly fix it.'' This observation aligns with the higher \textit{drives} attributes in ChatGPT-generated answers. However, ChatGPT answers do not convey risks as much as human answers do. This indicates that human answers on Stack Overflow often warn programmers of the side effects of solutions more than ChatGPT does.

For \textit{\textbf{informal}} attributes, human answers are highly informal and casual. On the contrary, ChatGPT answers are very formal and do not make use of swear words, netspeak, nonfluencies, or fillers. In our observation, we rarely saw ChatGPT using a casual conversation style. On the other hand, human answers often had words such as ``btw'', ``I guess'', etc. Human answers also contain higher \textit{\textbf{perceptual}} and \textit{\textbf{cognitive}} keywords than ChatGPT answers. According to the definitions of \textit{\textbf{perceptual}} and \textit{\textbf{cognitive}} attributes by LIWC (Section \ref{sec:linguistic}), this indicates that human answers portray more personal observations and insights from human programmers when answering the question.  

\begin{eqbox}[\textbf{Finding 5}]
 Compared to human answers, ChatGPT answers are more formal, express more analytic thinking, showcase more efforts towards achieving goals, and exhibit less negative emotion. 
\end{eqbox}

\subsection{RQ5: Sentiment Analysis}
 %We computed the percentage of positive, neutral, and negative labels for both ChatGPT and SO answers as predicted by the sentiment analysis model. 
 Our results show that, among the 2000 ChatGPT answers,  1707 (85.35\%) of them portray positive sentiment, 291 answers (14.55\%) portray neutral sentiment, and only 2 answers (0.1\%) portray negative sentiment. On the other hand, 1466 of the 2000 SO answers (73.30\%) portray positive sentiment, 513 answers (25.65\%) portray neutral, and 21 answers (1.05\%) portray negative sentiment. To assess the sentiment difference between ChatGPT and SO answers, we performed a McNemar-Bowker test on the sentiments. Since we have paired-nominal data, we opted for the McNemar-Bowker test for testing the goodness of fit when comparing the distribution of counts of each label. The results are statistically significant ($X^{2}=186.84, df=3, p<0.001$).  Our results show that for 13.90\% questions, ChatGPT answers portrayed positive sentiment while human answers portrayed neutral or negative sentiments. On the other hand, only 2 ChatGPT answers portrayed negative sentiment when the human answers were positive or neutral. Our result indicates that ChatGPT shows significantly more positive sentiment compared to human answers.

\begin{eqbox}[\textbf{Finding 6}]
ChatGPT answers portray significantly more positive sentiments compared to human answers on Stack Overflow. 
\end{eqbox}

%% file: sections/Result_userstudy.tex
\section{User Study Results}
\label{result:user}
\begin{figure*}[]
    \centering
    \includegraphics[width=0.60\textwidth]{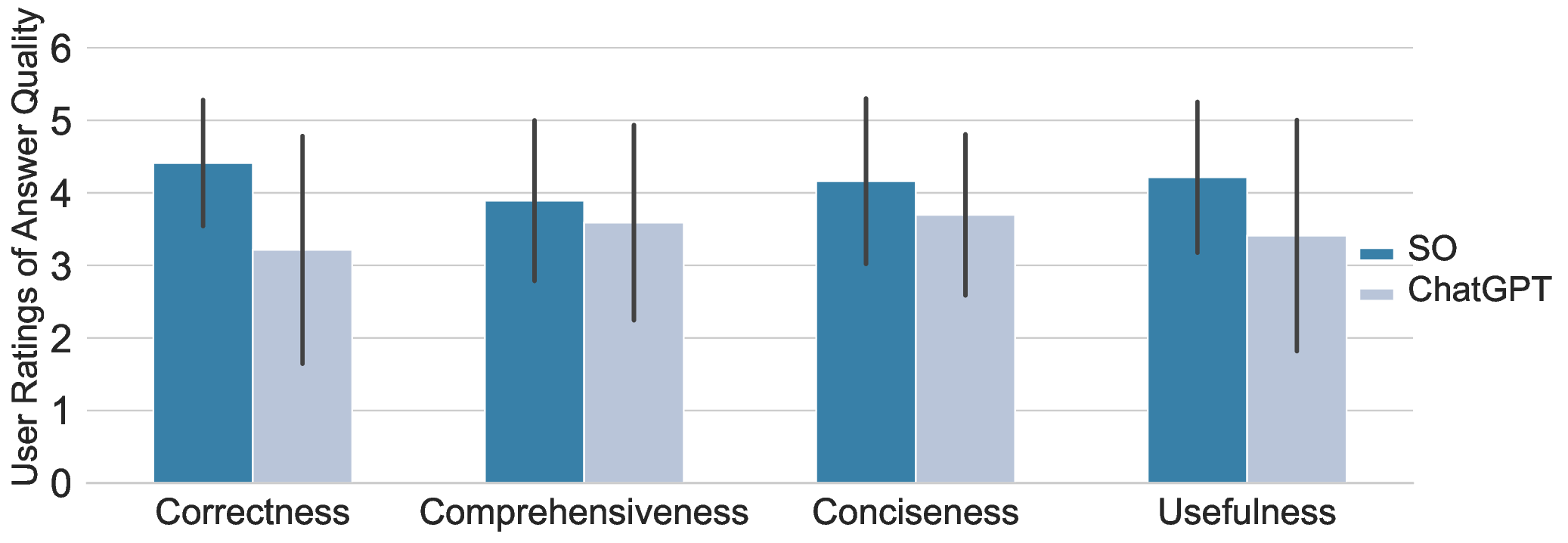}
    \vspace{-10pt}
    \caption{Quality of answers as rated by participants. Difference in \textit{Correctness, Conciseness}, and \textit{Usefulness} are statistically significant (Paired t-Test, p-value<0.05)}
    \label{fig:userrating}
    \vspace{-15pt}
\end{figure*}

We retrieved 56 pairs of ratings of ChatGPT answers and human answers as rated by 12 participants. Figure \ref{fig:userrating} presents the average ratings of the two kinds of answers in all four quality aspects. Overall, users found human answers to be more correct (mean rating human: 4.41, ChatGPT: 3.21, Welch’s t-test: p-value<0.001), more concise (human: 4.16, ChatGPT: 3.69, Welch’s t-test: p-value<0.05), and more useful (human: 4.21, ChatGPT: 3.42, Welch’s t-test: p-value<0.01).
For comprehensiveness, the average ratings are 3.89 and 3.98 for human answers and ChatGPT answers respectively. However, this result is not statistically significant.  

%To understand which factors play a role in user preference, identification of ChatGPT answers, and determining incorrectness, we conducted a thematic analysis on the interview transcripts. 
Additionally, our thematic analysis revealed five themes---\textit{Process of differentiating ChatGPT answers from human answers}, \textit{Heuristics of verifying correctness}, \textit{Reasons for incorrect determination}, \textit{Desired support}, and \textit{Factors that influence user preference}. %Codes among each theme are non-exclusive, meaning one transcript has multiple codes from each theme. 
Findings from our quantitative and thematic analysis for each of the research questions are described in the following subsections. 

% Moreover, we ran Pearson's Chi-Square test to evaluate the relationship between the incapability to identify incorrect answers and topic expertise. Our results show that if a participant has limited or no expertise in a topic, it is highly likely that they can not identify the incorrect answer (p-value  ). \todo{didn't hold to avoid}

% \vspace{-6pt}
\subsection{RQ6: Differentiating ChatGPT answers from human answers}
Our study results show that participants successfully identified which one is the machine-generated answer 80.75\% of the time and failed only 1\revision{9}.25\% of the time (Welch’s t-Test, p-value<0.001).

From thematic analysis, we identified the factors that participants found helpful to discern ChatGPT answers from human answers. 6 out of 12 participants reported the writing style of answers to be helpful in identifying the ChatGPT answer. Participant P5 mentioned, {\em``good grammar''}, and P8 mentioned, {\em``header, body, summary format''} to be contributing factors for identification. Two other factors are language style (e.g., casual or formal language, format) (10 out of 12 participants) and length (7 out of 12 participants). Additionally, 5 participants found unexpected or impossible errors as a helpful factor in identifying the machine-generated answers. Apart from these, tricks and insights that only experienced people can provide (5 out of 12 participants), and high entropy generation (1 out of 12 participants) were two other reported factors.  Our result suggests that most participants use language and writing styles, length, and the presence of abnormal errors to determine the source of an answer. 

\vspace{-4pt}
\begin{eqbox}[\textbf{Finding 7}]
Participants can correctly discern ChatGPT answers from human answers over 80\% of the time. They look for factors such as formal language, structured writing, answer length, or unusual errors to decide whether an answer is generated by ChatGPT. 
\end{eqbox}
% \vspace{-6pt}
\subsection{RQ7: Assessing Answer Correctness}
Our study result shows that users could successfully identify the incorrect answers only 60.66\% of the time and failed 39.34\% of the time (Welch’s t-test, p-value<0.05). 

When we asked users how they identified incorrect information in an answer, we received three types of responses. 10 out of 12 participants mentioned they read through the answer, tried to find any logical flaws, and tried to assess if the reasoning made sense. 7 participants mentioned they identified the terminology and concepts they were not familiar with and did a Google search, and read documentation to verify the solutions. And lastly, 4 out of 12 users mentioned that they compared the two answers and tried to understand which one made more sense to them. \revision{All of the aforementioned verification processes involved assessing the code or part of the answers in external IDEs.  All of our participants copied code or tested part of the solution from at least one answer into their local IDE for validation, 9 participants utilized some online code sandbox for validation (e.g., sandbox for HTML, CSS, JS), and 6 participants used the built-in code editor from tutorial sites such as W3Schools as a part of their assessment process. }

When a participant failed to correctly identify the incorrect answer, we asked them what could be the contributing factors. 7 out of 12 participants mentioned the logical and insightful explanations, and comprehensive and easy-to-read solutions generated by ChatGPT made them believe it to be correct. 6 participants mentioned lack of expertise to be the reason. However, we ran Pearson's Chi-Square test to evaluate the relationship between overlooking incorrect answers and topic expertise and found no significant relation between these two. P7 and P10 said ChatGPT's ability to mimic human answers made them trust the incorrect answers. 
% We believe the writing style, comprehensiveness, and linguistics features of ChatGPT answers impair users' ability to identify incorrect answers. \textbf{(RQ7)}

\begin{eqbox}[\textbf{Finding 8}]
 Users overlook incorrect information in ChatGPT answers 39.34\% of the time due to the comprehensive, well-articulated, and humanoid insights in ChatGPT answers. 
\end{eqbox}

Additionally, participants expressed their desire for tools and support that can help them verify the correctness. 10 out of 12 participants emphasized the necessity of verifying answers generated by ChatGPT before using it. Participants also suggested adding links to official documentation and supporting in-situ execution of generated code to ease the validation process. 

\subsection{RQ8: Factors for User Preference}
Participants preferred SO answers 65.18\% of the time. However, participants still preferred ChatGPT answers 34.82\% of the time (Welch’s t-test, p-value<0.01). Among the ChatGPT preferences, 77.27\% of the answers were incorrect. 

For \textit{factors that influence user preference}, 10 out of 12 participants mentioned correctness to be the main contributing factor for preference. 8 participants mentioned answer quality (e.g., conciseness, comprehensiveness) as contributing factors. 6 participants mentioned they put emphasis on how insightful and informative the answer is while preferring. 6 participants stated language style to be one of the factors, 2 of these 6 participants preferred the casual, spontaneous language style of human answer, while the other 4 preferred the well-structured and polite language of ChatGPT. P2 mentioned, {\em``It feels like it's trying to teach me something''}. 
% P10 stated the background and summary sentences at the beginning and end of the answers make them easy to read and understand.
Finally, 5 participants mentioned the format, look and feel (e.g., highlighting, color scheme) as contributing factors toward preference. 

\begin{eqbox}[\textbf{Finding 9}]
% \vspace{-5pt}
Participants preferred human answers from Stack Overflow more than ChatGPT answers (65.18\% of the time). Participants found human answers to be more correct, concise, and useful. %A few reasons for SO preferences are -- correctness, conciseness, casual and spontaneous language, etc.
% \vspace{-5pt}
\end{eqbox}

%% file: sections/Discussion.tex
\section{Discussion and Future Work}
%In this work, we have empirically studied ChatGPT answers to programming questions. While our results and findings give an overall understanding of the different characteristics of ChatGPT responses through the lens of manual and linguistics analysis, in this section we discuss the implications of these findings and future research directions and caution for stakeholders in this field. 
In this section, we discuss the implications of our findings and future directions to counter misinformation when using ChatGPT for programming.

\subsection{Why Do Users Prefer ChatGPT Responses?}

%\textbf{Users get tricked by the linguistic style of an answer.} 
Surprisingly, our user study shows that participants preferred ChatGPT answers 34.82\% of the time, though 77.27\% of these answers contained misinformation. Furthermore, we observed that participants overlooked a lot of misinformation in ChatGPT answers. Specifically, when ChatGPT answers are not readily verifiable (e.g., requiring execution in an IDE or needing to go through long documentation to validate), users often fail to identify the misinformation and underestimate the degree of incorrectness in the answer. %Surprisingly, even when the answer has an obvious error, 2 out of 12 participants still marked them as correct and preferred that answer. 
The follow-up semi-structured interviews revealed that the polite language, articulated and text-book style answers, and comprehensiveness are some of the main reasons that %, drives, and \todo{what is affiliation?}\new{it's a linguistic attribute in LIWC, this means ally, friendly conversation. For example, if a text has a high affiliation attribute it will give you a vibe of a more friendly conversation, or the person is answering as an ally.} affiliation 
made ChatGPT answers look more convincing, so the participants lowered their guard and overlooked some misinformation in ChatGPT answers. This finding is consistent with previous findings of user preferences over Stack Overflow (SO) posts. Prior work~\cite{nasehi2012makes, bazelli2013personality, calefato2015mining, asaduzzaman2013answering} shows that SO users preferred posts that contain illustrations, step-by-step instructions, multiple solutions, and positive sentiments. %Calefato et al.~\cite{calefato2015mining} listed positive affect (e.g., emotion, sentiment) as a success factor for SO answers and Asaduzzaman et al.~\cite{asaduzzaman2013answering} show that the presence of strong negative emotions discourages followup discussions in SO. Since our linguistic and sentiment analyses show that ChatGPT consistently portrays more positive sentiment than SO answers, our findings align with the previous work correlating the success of programming answers and positive sentiment. 
%Additionally, among the attributes for successful SO answers as listed by Nasehi et al.~\cite{nasehi2012makes}, step-by-step solutions, multiple solutions, and additional resources are some of the major attributes. 
Our linguistic analysis shows that ChatGPT answers possess many of these linguistic characteristics that SO users appreciate. 

Recently, there has been a decline in network traffic to the Stack Overflow website, which was attributed to the rise of ChatGPT~\cite{sodec}. Although our user study does not evaluate what encourages users to ask a question to ChatGPT rather than Stack Overflow in the first place, our findings point to some possible reasons. We believe the fact that users can avoid the embarrassment of posting online and the risk of receiving negative comments but still receive seemingly high-quality answers in a timely manner can be some contributors. 
\revision{Moreover, the interactive feature of ChatGPT makes it easier for users to change prompts and interactively work with the language model to make it generate desired or optimal answers. Using interactivity to rectify errors can be another contribution to ChatGPT's popularity among programmers.}

% Hence it is crucial to communicate the level of correctness to users. 

%\textbf{Users appreciate comprehensiveness and politeness.} Our user study reveals that \todo{this claim is not correct since users only prefer ChatGPT answers 30%+ times but prefer SO answers 60\%+ times. So it is not true to say that "users often prefer ChatGPT answers over SO answers"}users often prefer ChatGPT answers over SO answers irrespective of their correctness and based on how polite, positive, comprehensive, or well-structured the answers are. 

\subsection{Where Do Errors in ChatGPT Answers Emerge from? }
It is evident from our results that ChatGPT produces incorrect answers more than half of the time. Our observation sheds light on three main reasons for these errors. 

\textbf{Lack of Understanding for Some Programming Concepts.} 
First, 54\% of the time, errors are made due to ChatGPT not understanding the concepts mentioned in a question. For example, we found a JavaScript question about a website not showing the File Upload option~\cite{soexamplearea}. Clearly, it is an issue with the front end and User Interface (UI), since the question mentioned ``the file upload area not working'' and provided JavaScript and HTML code snippets. %Any programmer with basic HTML and JavaScript knowledge will understand from a quick glance that the posted problem and the code to be debugged is about the UI.
In this context, ``the file upload area'' refers to the UI widget to upload a file, rather than the action of uploading a file. ChatGPT did not get this and answered a handful of irrelevant solutions, such as how the file path needs to be set, how to locate the file in your machine, CORS issues, etc. By contrast, the human-written answer suggests adding an appropriate \textit{id} to the file input field in the HTML code. 
These types of misunderstanding issues contribute to the high number of \textit{Conceptual} errors. 

\textbf{Limited Capability to Understand and Reason Program Semantics.} Our manual analysis reveals that while most of the code examples (98\%) generated by ChatGPT are syntactically correct, many of them contain incorrect logic (48\%) or incorrect API usage (39\%). We suspect this is largely due to ChatGPT's limited capability to understand and reason program semantics. In many cases, ChatGPT makes obvious programming mistakes that human programmers barely make. For example, ChatGPT may generate a loop ending condition that is always true or false, e.g. \texttt{while(i<0 \&\& i>10)}. Furthermore, the content generation process in ChatGPT is essentially an auto-regressive decoding process guided by the probability distribution at each token prediction step. Thus, ChatGPT cannot foresee the potential outcome or execution result of the generated code. For example, we observed that ChatGPT generated a code example that keeps decreasing a variable in a for loop and eventually leads to a division-by-zero exception in the end. ChatGPT seems unable to understand the consequences or side effects of some code operations and expressions.

\textbf{Missing or Incorrect Attention to a Programming Question.} Since questions asked in SO are long human-written questions with many components involved, ChatGPT often focuses on the wrong part of the question or gives high-level solutions without fully understanding the minute details of a problem. 
For example, we found an instance where the SO question asked about differences between public, private, and protected access modifiers in Java. However, ChatGPT only focused on the part ``access modifiers'' ignoring the ``difference'' part in the question. Therefore, it gives an extremely verbose response that contains the definitions of encapsulation, inheritance, etc., which is not useful in terms of identifying the differences originally asked for.

% \todo{Please comment out the following reason since "hallucination" is a phenomenon. Not a reason.}
% Third, although the presence of \textit{Factual} error is lower than \textit{Conceptual} error but still composes a large portion (36\%) of incorrectness. We believe these \textit{Factual} errors stem from hallucinations where ChatGPT provides fabricated information. 

% \todo{Moving the following discussion to future actions for academic researchers. Also condense it since i saw many repetitions to the discussion above.}

\subsection{What Is at Stake and What Does the Future Hold?}

\textbf{Impact on the Software Industry and Society.} We believe that the large number of seemingly correct ChatGPT answers pose high risks to programming practices since they can easily trick programmers into thinking they are correct, especially when programmers lack the expertise or means to readily verify the correctness. As AI Chain frameworks are getting increasingly popular, it becomes even riskier when the ChatGPT answers are automatically integrated into downstream AI components with no human involvement and validation. The misinformation will propagate along the AI chain and may have devastating effects on downstream tasks. In the long term, this could jeopardize the quality and robustness of software and cyberinfrastructure in our society, since the misinformation in these answers may lead to suboptimal design decisions and software defects. The repercussions can potentially affect other societal factors, including the safety, security, and trust of the general population.

%Furthermore, the existence of a large number of errors in answers to programming questions raises concerns surrounding the risks that might reach beyond the programmers and software and affect the general population. Incorrect and insecure code have the risk of seeping into commercial software products and affecting all parties involved in the software manufacturing chain including the manufacturer, vendor, consumer, etc. Apart from economic aspects, other factors such as user safety, security, and trust are prone to the threats posed by misinformation.  

\textbf{Impact on STEM Education.} Many STEM fields, beyond Computer Science, require students to learn basic programming. Students using ChatGPT for learning materials may be misled into learning incorrect concepts and information. This may even harm the grades or reputation of students. We believe identifying and verifying errors in programming answers require as much expertise as learning and writing code. Hence, learning through the wrong materials has the potential to create a chain of misinformation where the veracity assessment of students and learners will be compromised in the long term. %Previous work~\cite{zhou2023synthetic, kreps2022all,heaven2022meta} show the existence and effects of misinformation in several contexts including but not limited to fake news, fake scientific materials, etc. Buchanan et al.~\cite{buchanan2021truth} shows the alarming capability of GPT-3 to produce disinformation at large scale. Our findings add to the previous work in the context of software engineering and programming. We believe that ChatGPT shows a similar capability of producing insecure code and incorrect code at a large scale. Therefore, it is of vital importance that misinformation in codes and programs should be identified, monitored, and mitigated.  

\textbf{The Silver Lining.} While our manual analysis reveals 52\% of the answers are incorrect, 48\% answers are completely correct \revision{(i.e., no statements in those answers annotated with factual, conceptual, code, or terminological errors)}, which by no means is an insignificant number. Compared with Stack Overflow, ChatGPT can give immediate answers to users' questions, significantly saving the time and effort of users.  Thus, conversational chatbots such as ChatGPT may be considered more convenient than Q\&A forums. Programmers of all levels, including students and professional developers, may find it easy and less time-consuming to ask basic programming questions instead of going to instructors, mentors, or even posting on traditional Q\&A platforms.

Hence, along with trying to rectify the error and mitigate the risks, steps should be taken to create awareness and adopt new strategies and policies to address the risks associated with incorrect information generated by ChatGPT.

\subsection{What Further Actions are Needed to Address Misinformation in ChatGPT?}

\subsubsection{Limitations of Existing Approaches} 

%\textbf{Prompt Engineering is handy but a more sustainable solution is needed.} 
Although approaches have been proposed to mitigate hallucinations from \textit{LLM}s~\cite{dong2022calibrating,peng2023check}, they are only applicable to fixing \textit{Factual} errors. Since the root of \textit{Conceptual} errors is not hallucinations but rather a lack of understanding of programming concepts and incapability to reason program semantics, existing approaches for hallucination may not be effective in mitigating conceptual errors.

Most of the existing methods to help \textit{LLM}s understand and reason rely on \textit{Prompt Engineering}. While \textit{Prompt Engineering} can be helpful in probing ChatGPT to understand a problem to some extent~\cite{zhou2022large,strobelt2022interactive}, they are still insufficient when it comes to injecting reasoning into \textit{LLM}s to solve special cases. Moreover, \textit{Prompt Engineering} is not a sustainable solution and the responsibility largely falls on users.

\revision{Furthermore, ChatGPT provides different answers even when prompted with the same questions. This makes the verification process even harder since users cannot deterministically identify the prompts that will always result in correct or optimal solutions. Although lowering the temperature value can help in achieving consistent answers for the same prompts, lower temperature often reduces the quality of answers generated by \textit{LLM}s. Thus, this variability adds another dimension to the challenges already posed by \textit{Prompt Engineering}.}
Additionally, \textit{Prompt Engineering} implies that to make ChatGPT give the right answer, users need to ask the right question. Thus, overly relying on \textit{Prompt Engineering} to make ChatGPT produce the correct answer shifts the responsibility for AI errors to humans. 
% Moreover, as pointed out by previous work~\cite{bender2020climbing,bender2021dangers}, training signals should contain some form of grounding signals other than just linguistic forms. We encourage future research in the direction of developing training data that goes beyond just codes. Although in many current cases codes in the training data are accompanied by some test data that provides grounding information to some extent, as pointed out by~\cite{bender2020climbing}, better interaction and grounding data are needed in training signals to aid in NLU. 
Hence, we urge that instead of temporary patches such as changing prompts that also make humans somewhat responsible for the errors made by ChatGPT, it is essential to understand the sources and factors of conceptual errors in order to develop sustainable and special-purpose solutions to fix them.

\subsubsection{Communicating the level of incorrectness is necessary.} The user interface of ChatGPT includes a one-line warning---``\textit{ChatGPT may produce inaccurate information about people, places, or facts.}'' However, we believe such a generic warning is insufficient. Each answer should be accompanied by a level of incorrectness and uncertainty in the answer. Moreover, our observations indicate that not all answers have an equal amount of incorrectness---some answers have the majority of parts marked as incorrect, whereas some answers have only a few lines marked as incorrect. Since each incorrect answer differs in the severity of incorrectness, it is vitally important to provide users with the level of incorrectness for each answer. A recent study shows that an \textit{LLM} may know when it is lying ~\cite{azaria2023internal}, which can be leveraged to warn users about the potential errors made by LLMs. %, but does \textit{LLM} know when it is speculating? And how can we communicate the level of speculation? 
%Previous work~\cite{agarwal2020quality} has shown that communicating the confidence level does not always help programmers with better interpretability of the quality of auto-generated code. 
However, recent  studies~\cite{agarwal2020quality, vasconcelos2023generation} also show that only rendering the confidence level is not sufficient to help programmers understand the uncertainty and risks in the generated code. Thus, it is necessary to investigate more effective communication and visualization methods for model uncertainty in programming tasks. 
%Therefore, it is imperative to investigate how to communicate the level of incorrectness of the programming answers and code solutions. Moreover, human factor research on communicating the level of incorrectness in codes or programming answers in a way that users understand is another important direction worth exploring. 

Moreover, for software companies, it is worthwhile to invest in more awareness campaigns and training for software developers. Special training is necessary for software developers so that they can monitor the code bases, readily verify errors in ChatGPT answers, and perform more testing to safeguard errors from sneaking into their codebases. In particular, software developers should be advised to use ChatGPT with more caution and scrutiny for high-stake code blocks and programming tasks. 

%\subsubsection{Awareness, Tools, and Monitoring are Essential \qquad \qquad \qquad \qquad \qquad \qquad \qquad \qquad  \qquad \qquad \qquad \qquad \qquad \qquad \qquad \qquad} 
\subsubsection{More rigorous code reviews and testing are needed.} 
%\textbf{Rigorous and special purpose software testing is needed.} 
Software companies should enforce more rigorous code reviews and software testing methods to source code that is produced with the facilitation of ChatGPT and other AI technologies. Since ChatGPT may make programming mistakes that human programmers barely make, it is important to adapt traditional methods to account for the types of programming mistakes generated by ChatGPT or other \textit{LLM}s. Additionally, it is necessary to have continuous testing and security checking so that incorrect or insecure code can not seep into any part of the software products. 
\revision{Moreover, ChatGPT can be integrated into the testing pipeline as ChatGPT can potentially generate test cases on the fly. Hence, encouraging the integration of testing during the generation process can limit the risk of programming mistakes made by ChatGPT.}

\subsubsection{Future actions for academics and researchers} 
% We argue it is important to distinguish between \textit{Conceptual} and \textit{Factual} errors because their root cause is distinctive in nature and requires different actions to address them. 
Bender and Koll- er~\cite{bender2020climbing} show that any \textit{LLM}s trained only on the form of language can not fully reach the human level of understanding. They argued that to aid \textit{LLM}s in performing natural language understanding, it is imperative to have information in the training data that goes beyond just the form of language, e.g., code paired with several input and correlated output, edge cases, etc. Furthermore, Bender and Gebru et al.~\cite{bender2021dangers} argue that increasing the size of language models is not a solution to achieving natural language understanding. 
We believe one of the main reasons behind the large number of conceptual errors can be attributed to ChatGPT's limitation in performing natural language understanding. Moreover, although existing work~\cite{xiong2017deeppath,chen2020review} shows the challenges and limitations of reasoning in \textit{LLM}s and presents Knowledge Graphs as a powerful method to aid in reasoning, our results highlight the limitation in reasoning when it comes to programming answers or code solutions. Therefore, we urge the attention of the research community for rigorous investigation and mitigation methods to improve the reasoning and understanding capability of \textit{LLM}s, especially in the field of programming. 
% And finally, since the nature of the root cause of \textit{Conceptual} errors is utterly different than hallucination, special identification and mitigation methods need to be adopted for \textit{Conceptual} errors while reducing hallucination at the same time. 

\subsubsection{Implications for code reviewers and teaching staff in STEM classrooms.} Previous work~\cite{jiang2018linguistic,rashkin2017truth} shows that linguistic features can be used as a mechanism to identify misinformation and AI-generated content. Our results show that ChatGPT answers have a very distinct linguistic structure and communication style when answering programming questions. We believe identifying these distinct linguistic features is essential in situations where users need to differentiate between human and machine-generated answers. For example, in CS classrooms, there is an increasing concern that students are using ChatGPT to solve homework assignments, which is an impediment to learning. Traditional plagiarism tools used in academia often cannot detect ChatGPT-generated answers. By having general knowledge of common language styles of ChatGPT answers (e.g., verbosity, formal language, title-body-summary structure, etc.), teaching staff can be more aware of what to look for. Moreover, plagiarism tools, both AI and non-AI, should incorporate unique linguistic characteristics as factors to classify plagiarised documents. Furthermore, as discussed in the previous subsections, it is imperative that code reviewers adopt new techniques and tools to take extra precautions so that incorrect and insecure code does not seep into software products. Incorporating the linguistics style of ChatGPT responses while creating these tools and training code reviewers to make them aware of unique linguistic markers can help the software industry install additional safeguards against incorrect code. 

\subsubsection{New pedagogical methods are necessary.} Apart from the software industry, faculty and teaching staff in the educational institute should also make the students aware of the potential risks that come with seemingly correct ChatGPT answers. Moreover, new pedagogical methods should be adopted to incorporate ChatGPT into the curriculum to utilize the incorrectness as a learning tool. For example, in a beginner Python class, students can be given multiple wrong programs generated by ChatGPT and asked to identify the errors in each program. This type of activity can render learning and create awareness at the same time.

% \subsubsection{Responsible AI and Protecting Human Interest \qquad \qquad \qquad \qquad \qquad \qquad \qquad \qquad  \qquad \qquad \qquad \qquad \qquad \qquad \qquad \qquad \qquad} 

% \textbf{Automated tool is helpful but human intervention is essential.}
% While we suggest the necessity of having an automated tool for detecting incorrect or insecure codes, we also believe that depending solely on these detection tools or asking ChatGPT to decide if a response is GPT generated, can be harmful in many cases. Since ChatGPT does not provide a confidence level, and these automated tools might not have 100\% accuracy, it is highly probable that these methods will flag correct codes as incorrect codes and non-plagiarized answers as plagiarized answers.  When AI is trusted with the decision to decide on AI-generated materials, it creates a cascade of misinformation, affecting the goodwill of stakeholders at each level. These situations have the potential to have devastating effects on the students' or programmers' morale. We urge that instead of only depending on automated detection tools, stakeholders such as teachers, TAs, and code reviewers should have generic training on the linguistic differences between human and machine-generated programming answers and should monitor the detection tools.

\subsubsection{Separation of accountability.} New policies should be made to separate and distinguish the role of humans and \textit{LLM}s when \textit{LLM}s generate misinformation. As discussed previously, depending on solutions such as \textit{Prompt Engineering} shifts the accountability of misinformation to humans. Furthermore, when AI is involved in the step of decision-making and manufacturing software products, ethical questions such as who will be held accountable for AI's errors come to light~\cite{deshpande2022responsible}. 
Previous work~\cite{binns2018s,tahaei2023human} on responsible AI also highlights the need for accountability of AI systems that are grounded in human rights and ethics. Hence, strict policies should be created to maintain the separation of accountability to protect humans from false accusations and preserve the interests of impacted stakeholders by ensuring responsible use. We believe this work will encourage further research for the informed design of responsible conversational chatbots and for careful policy-making to preserve the rights of stakeholders. 

% We would like to conclude this discussion with the statement that, AI is most effective when supervised by humans. Therefore, we call for the responsible use of ChatGPT to increase human-AI productivity. 

%% file: sections/Threat.tex
% \vspace{-5pt}
\section{Limitations}

One limitation of this work is the subjective nature of the manual analysis. We tried to address this limitation by recruiting multiple labelers, constantly measuring the agreement level among labelers, and adopting an iterative analysis procedure with extensive discussions. Moreover, our user study has limitations concerning other factors such as sample size and participants' own biases. %Although 12 is a small sample size, each participant provided us with 5 data points on average (56 in total). We believe this addresses the limitation due to the small number of participants to some extent. Furthermore, 
%Since our user study is designed to be a qualitative study to complement, the manual and linguistic analysis, we believe the statistically significant results from the sample size and collected data points successfully back up the findings from manual and linguistic analysis. 
To reduce participants' biases against human or ChatGPT answers, we anonymized the source of the answers during the study and standardized the visual style and format of the answers, e.g., using the same font size, type, code style, etc. %For thematic analysis, we followed several iterations to make a comprehensive list of codes (21) and 5 themes to capture minute details from user feedback and reduce subjective human bias. Apart from these, our user study is a controlled lab study and does not assess user behavior in-the-wild such as what motivates a user to use ChatGPT, or how users change their prompts and questions after identifying incorrect answers, etc. Hence, we invite future research to understand user behavior and engagement with ChatGPT outside of a controlled lab environment. 
% \revision{Verifying solutions with longer code was another challenge participants faced in the user study. ...}

Additionally, this work has used the free version of ChatGPT (GPT-3.5) for acquiring the ChatGPT responses for the manual analysis. Hence, one might argue that the results are not generalizable for ChatGPT since the new GPT-4 (released on March 2023) can perform differently. 
\revision{To understand how differently GPT-4 performs compared to GPT-3.5, we conducted a small analysis on 21 randomly selected SO questions where GPT-3.5 gave incorrect answers. \footnote{The annotations for GPT-4 are added in the repository \url{https://github.com/SamiaKabir/ChatGPT-Answers-to-SO-questions/blob/main/ChatGPT\%20answers\%20to\%20SO\%20questions/Labeler1/Annotations_GPT-4.docx}} Our analysis shows that, among these 21 questions, GPT-4 could answer only 6 questions correctly, and 15 questions were still answered incorrectly. Moreover, the types of errors introduced by GPT-4 follow the same pattern as GPT-3.5. This tells us that, although GPT-4 performs slightly better than GPT -3.5 (e.g., rectified error in 6 answers), the rate of inaccuracy is still high with similar types of errors. } 
Moreover, this new ChatGPT (also known as ChatGPT plus) is a paid version (\$20 per month). Since the target population of this research is not only industry developers but also programmers of all levels, including students and freelancers around the world, the free version of ChatGPT has significantly more users than the paid version which only the privileged population can access. Moreover, \$20 per month has a considerably high monetary value for many countries. Hence, for this study, we used the free version (GPT-3.5) so that the results benefit the majority of our target populations. We acknowledge that other \textit{LLM}s can perform differently and we encourage future research to empirically study programming answers generated by other \textit{LLM}s.

\revision{
Another limitation lies in the prompting strategy adopted by our study. In this work, we did not account for the interactive nature of ChatGPT. In practice, if the initial ChatGPT answer is not satisfactory, programmers can refine their initial prompt or ask follow-up questions to get new answers. We did not consider this, since it required designing specific follow-up questions or prompt refinements for each question under analysis. Furthermore, such interaction is not guaranteed to generate better and more correct answers. Thus, it may require multiple rounds of interaction to improve the answer. This would significantly increase the analysis effort and limit our capability to analyze many different kinds of questions in this study. As a result, we restrict the project scope to only analyze the initial answers generated by ChatGPT. To address this limitation, future work could conduct a small-scale but more focused analysis to investigate how interactivity impacts the correctness of ChatGPT answers.} 

%Although changing prompts and interacting with ChatGPT can generate different responses, there is no guarantee that newer prompts derived from interactivity will result in correct answers. Moreover, there is no valid study that gives clear instructions on what type of prompts are effective given different problem scenarios. Furthermore, since the design of the prompt is highly dependent on the problem itself and also varies from person to person, reaching an agreement level for prompt engineering is even more challenging without any established guidelines or studies to follow. Hence, interactively identifying issues in ChatGPT responses and formulating appropriate prompts are parts of a bigger problem and are out of the scope of this work. Future work should emphasize conducting a systematic investigation into how different prompting strategies and tips influence the correctness of ChatGPT answers to different kinds of programming questions. 
\revision{In this work, we reused the original SO question as the prompt, since the original SO question represents how a programmer may ask the question in a natural conversation. This can be improved with more advanced prompting templates and tricks. However, the design of the prompt is highly dependent on the problem itself and also varies from person to person. Reaching an agreement level for prompt engineering is even more challenging without any established guidelines or studies to follow. To address this limitation, future work could conduct a systematic investigation into how different prompting strategies and tips influence the correctness of ChatGPT answers to different kinds of programming questions.} 

\revision{ChatGPT is inherently stochastic. The same prompt may generate different answers with a moderate temperature setting of 0.8. To account for this, one needs to run ChatGPT multiple times with the same prompt for each programming question, manually analyze all answers, and measure the average correctness. If we run ChatGPT 5 times for each question, our analysis effort would be increased by five times and we would not be able to do the study at a satisfiable scale and comprehensiveness level. Thus, we chose to only consider the initial answer generated by ChatGPT.  %To mitigate this, we adapted a zero-shot approach by prompting ChatGPT just once with questions from SO that are already understood and answered by humans. Moreover, the original SO question also represents how a programmer may ask the question in a natural conversation which makes a fair comparison between ChatGPT and SO. 
%Future work should focus on tailoring guidelines for prompt engineering that can account for interactivity and at the same time help with manual verification and large-scale analysis as this work. Furthermore, we encourage more research on determining methods for qualitative analysis that can account for variability among solutions for the same problem based on level and type of interactivity.
}

Finally, we acknowledge that despite our efforts to mitigate the potential issues, some level of human bias and the generalizability limitation still persist. 
Nonetheless, we hope this study will foster new research in the direction of identification, understanding, rectification, and risk mitigation of errors in \textit{LLM}s for better human-AI collaboration. 

% Despite our best efforts, some level of bias and threats to internal validity still persist. 

% \textbf{External Validity.} 

% Another limitation is the generalizability of the findings of our analyses. To address this, we collected SO answers across diverse categories and topics. Furthermore, we recruited participants with different levels of expertise in programming and in individual topics. 

%% file: sections/Conclusion.tex
\section{Conclusion}

In this paper, we empirically studied the characteristics of ChatGPT answers to SO questions through a combination of manual analysis, linguistic analysis, and user study. Our manual analysis shows that ChatGPT produces incorrect answers more than 50\% of the time. Moreover, ChatGPT suffers from other quality issues such as verbosity, inconsistency, etc. Results of the in-depth manual analysis also point towards a large number of conceptual and logical errors in ChatGPT answers. Additionally, our linguistic analysis results show that ChatGPT answers are very formal, and rarely portray negative sentiments. Although our user study shows higher user preference and quality rating for human answers, users make occasional mistakes by preferring incorrect ChatGPT answers based on ChatGPT's articulated language styles, as well as seemingly correct logic that is presented with positive assertions. Since ChatGPT produces a large number of incorrect answers, our results emphasize the necessity of caution and awareness regarding the usage of ChatGPT answers in programming tasks. This work also seeks to encourage further research in identifying and mitigating different types of conceptual and factual errors. Finally, we expect this work will foster more research on transparency and communication of incorrectness in machine-generated answers, especially in the context of programming.

%% file: appendix.tex
% \newpage
\clearpage
\appendix
% \newpage
\section{\textbf{EXAMPLE OF INCORRECT ChatGPT RESPONSE}}

\textbf{\colorbox{lightgray}{ \qquad \qquad \qquad \qquad  Stack Overflow Question \qquad \qquad \qquad \qquad}} \newline
\large{\textbf{How to end program in a do-while (C++)}}

\small{Here is my code. I am trying to get the entire program to end if it goes into the second if statement inside the do-while loop. But every time I run it, it crashes. I am not sure what I am doing wrong.}

\begin{minted}[fontsize=\footnotesize]{C++}
#include <iostream>
using namespace std;

int main() {
    int myData[10];
    for(int i=0;i<10;i++){
        myData[i] = 1;
        cout<<myData[i];
    }
    
    do{
        int i;
        cout<<endl<<"Input index: ";
        cin>> i;
        
        int v;
        cout<<endl<<"Input value: ";
        cin>>v;
        
        if(i>=0||i<10){
            myData[i]=v;
            for(int i=0;i<10;i++){
                cout<<myData[i]<<" ";
            }
        }
   
        if (i<0||i>=10){
            cout<<"Index out of range. Exit.";
            return 0;
        }
        
    }while(1);
} 
\end{minted}

\noindent \textbf{\colorbox{lightgray}{\qquad \qquad \qquad \qquad \qquad  ChatGPT Answer \qquad \qquad \qquad \qquad \qquad \qquad}} \newline
\small{The problem with your code is that the second if statement inside the do-while loop will always be executed because you are checking if i is less than 0 or greater than or equal to 10, but not both. In the case where i is equal to 10, both conditions will be true and the program will print ``Index out of range. Exit.'' and then end. 
To fix this, you should change the if statement to use the logical and operator (\&\&) instead of the logical or operator (||):}
% \vspace{20pt}
\newline
\textbf{C Code}
\begin{minted}[fontsize=\footnotesize]{C++}
if (i <0 && i >=10) 
{
    court << "Index out of range. Exit."; << return 0;
}
\end{minted}
With this change, the second if statement will only be executed if i is less than 0 and greater than or equal to 10, which is not possible.
\newline

\noindent \textbf{\colorbox{lightgray}{ \qquad \qquad \qquad \qquad \quad Stack Overflow Answer \qquad \qquad \qquad \qquad \qquad }} \newline
\begin{minted}[fontsize=\footnotesize]{C++}
if (i >= 0 || i < 10) {
\end{minted}
\small {Think about which numbers are either greater than zero or less than ten. I'm sure you realise that is true of all numbers. What you meant to write is}
\begin{minted}[fontsize=\footnotesize]{C++}
if (i >= 0 && i < 10) {
\end{minted}
This explains your crash, you are accessing the myData array with an index that is outside the array bounds.
\newline
It's very common for beginners to get \&\& and || confused, especially where there is negation involved as well.

\newpage
\section{\textbf{Example of verbose chatgpt response}}

\textbf{\colorbox{lightgray}{ \qquad \qquad \qquad \qquad \quad Stack Overflow Question \qquad \qquad \qquad \qquad \quad }} \newline
\large{\textbf{JavaScript check if variable exists (is defined/initialized)}}

\small{Which method of checking if a variable has been initialized is better/correct? (Assuming the variable could hold anything (string, int, object, function, etc.))}

\begin{minted}[fontsize=\footnotesize]{JavaScript}
if (elem) { 
\end{minted}

\small{or}

\begin{minted}[fontsize=\footnotesize]{JavaScript}
if (typeof elem !== 'undefined') { 
\end{minted}

\small{or}
\begin{minted}[fontsize=\footnotesize]{JavaScript}
if (elem != null) { 
\end{minted}

\vspace{10pt}

\noindent\textbf{\colorbox{lightgray}{\qquad \qquad \qquad \qquad \qquad  ChatGPT Answer \qquad \qquad \qquad \qquad \qquad \qquad}} \newline

The method of checking if a variable has been initialized depends on the specific use case and the expected value of the variable. Here's an overview of each method: 

if (elem) checks if elem is ``truthy'', meaning that it has a value that evaluates to true in a boolean context. This includes all non-falsy values, such as numbers (except 0), strings, and objects. However, this method will also return true for empty strings, zero, NaN, null, and undefined, which may not be desirable in all cases. 

if (typeof elem !== 'undefined') checks if elem has been defined. This method is more precise than the first one, as it only returns true if elem has been declared in the current scope, even if it has been assigned a falsy value. However, it can be more verbose and may not be necessary in all cases. 

if (elem != null) checks if elem is not null or undefined. This method is useful if you want to exclude null and undefined values, but may return true for other falsy values such as empty strings and zero. 

In general, it's a good practice to be explicit about the expected value of a variable and use the appropriate method to check for it. For example, if you're checking for the existence of an object property, you may want to use if ('propertyName' in object) to avoid potential errors with falsy values. 
\newline

\noindent \textbf{\colorbox{lightgray}{ \qquad \qquad \qquad \qquad \qquad Stack Overflow Answer \qquad \qquad \qquad \qquad \quad }} \newline

\noindent You want the \underline{\textcolor{blue}{typeof}} operator. Specifically:

\begin{minted}[fontsize=\footnotesize]{JavaScript}
if (typeof variable !== 'undefined') {
    // the variable is defined
} 
\end{minted}